\documentclass{aastex63}

\usepackage{amsmath}
\usepackage{braket}
\usepackage{bm}
\usepackage{amsmath,amssymb}
\usepackage{subfigure}
\usepackage{appendix}
\usepackage{capt-of}
\usepackage{booktabs}
\usepackage{varwidth}
\usepackage{graphicx}
\usepackage{paralist}
\usepackage{lineno}
\usepackage{xcolor}
\usepackage[normalem]{ulem}

\newcommand{\mode}[2]{{}_{#1}\mathrm{S}_{#2}\,}

\newcommand{\wigfull}[6]{\bigg(\begin{smallmatrix} #1 & #2 & #3 \\#4 & #5 & #6 \end{smallmatrix}\bigg)\,}

\newcommand{\cP}{\mathcal{P}^{(\ell)}}

\newcommand{\cA}{\ensuremath{\mathcal{A}}}

\newcommand{\bfxi}{\mbox{\boldmath $\bf \xi$}}

\newcommand{\sfZ}{\mbox{${\sf Z}$}}

\newcommand{\bfnabla}{\mbox{\boldmath $\bf \nabla$}}

\def\rmd{{\mathrm{d}}}

\newcommand{\tdot}{\,.\hspace{-0.98 mm}\raise.6ex\hbox{.}
                   \hspace{-0.98 mm}\raise1.2ex\hbox{.}\,}

\def\bfv{{\mathbf{v}}}

\submitjournal{\apjs}
\received{2022 July 18} 
\revised{2023 March 18}
\accepted{2023 March 23}

\shorttitle{Rotation inversions from QDPT}
\shortauthors{Bharati Das and Kashyap et al.}

\begin{document}
\title{Minuscule corrections to near-surface solar internal rotation using mode-coupling}

\correspondingauthor{Srijan Bharati Das}\leavevmode\\
\email{sbdas@princeton.edu}

\thanks{Both authors have contributed equally to this study.}
\author[0000-0003-0896-7972]{${}^*$Srijan Bharati Das}
\affiliation{Department of Geosciences \\
Princeton University \\
Princeton, New Jersey, USA}

\author[0000-0001-5443-5729]{${}^*$Samarth G. Kashyap}
\affil{Department of Astronomy and Astrophysics \\
Tata Institute of Fundamental Research \\
Mumbai, India}

\author[0000-0001-7104-0104]{Deniz Oktay}
\affil{Department of Computer Science \\
Princeton University \\
Princeton, New Jersey, USA}

\author[0000-0003-2896-1471]{Shravan M. Hanasoge}
\affil{Department of Astronomy and Astrophysics \\
Tata Institute of Fundamental Research \\
Mumbai, India}

\author[0000-0002-2742-8299]{Jeroen Tromp}
\affiliation{Department of Geosciences \\
and Program in Applied \& Computational Mathematics \\
Princeton University \\
Princeton, New Jersey, USA}



\begin{abstract}
The observed solar oscillation spectrum is influenced by internal perturbations
such as flows and structural asphericities. These features induce splitting of
characteristic frequencies and distort the resonant-mode eigenfunctions. Global axisymmertric flow ---
differential rotation --- is a very prominent perturbation. 
Tightly constrained rotation profiles as a function of latitude 
and radius are products of established helioseismic pipelines 
that use observed Dopplergrams to generate frequency-splitting 
measurements at high precision.  However, the inference 
of rotation using frequency-splittings do not consider
the effect of mode-coupling.
This approximation worsens for 
high-angular-degree modes, as they
become increasingly proximal in frequency. Since modes with high angular
degrees probe the near-surface layers of the Sun, inversions considering
coupled modes could potentially lead to more accurate estimates of rotation very close to the surface. In order to investigate if this is indeed the case, we perform inversions for solar differential rotation, considering coupling of modes for angular degrees
$160 \leq \ell \leq 300$ in the surface gravity $f$-branch and first-overtone
$p$ modes. In keeping with the character of mode coupling, we carry out a non-linear inversion using an eigenvalue solver. Differences in inverted profiles for frequency splitting measurements from MDI and HMI are
compared and discussed. We find that corrections to the near-surface differential rotation profile, when accounting for mode-coupling effects, are smaller than 0.003 nHz and hence are insignificant. These minuscule corrections are found to be correlated with the solar cycle.
We also present corrections to even-order splitting
coefficients, which could consequently impact inversions for structure 
and magnetic fields.
\end{abstract}




\section{Introduction} \label{sec:intro}

Helioseismology has enabled high-precision measurements of
solar internal rotation. Turbulence in the convection zone excites modes of oscillation which propagate through the solar interior and are sensitive to the prevalent structure and flows. The extent to which these modes of oscillation, called the solar normal modes, are coupled depends on their proximity in frequency. 
Rotation breaks spherical symmetry,
resulting in prograde modes with higher frequencies and retrograde
modes with lower frequencies than the corresponding eigenfrequencies predicted in static, non-rotating standard models \citep[such as Model-S as defined by][]{jcd}. This is called rotational frequency splitting.

Degenerate perturbation theory \citep[DPT;][]{lavely92} 
has traditionally been employed for estimating rotation, where coupling between distinctly different modes is ignored. DPT has been employed for inversions using helioseismic data from 
BBSO \citep{Libbrecht-1989-ApJ, Brown-1989-ApJ}, 
GONG \citep[e.g.,][]{Thompson-1996-Sci}, 
MDI \citep[e.g.,][]{Kosovichev-1997-SoPh, Schou-1998-ApJ} and
HMI \citep[e.g.,][]{larson18}. 
Differential rotation (DR) is an extensively studied feature in the Sun. It is beyond the scope of this paper to discuss the many important studies on DR, and the reader is referred to \cite{howe08_review} for a comprehensive review of all notable contributions. Of particular importance, from a methodological perspective, is \cite{Schou-1998-ApJ}, which compared inversions via seven different methods and identified various robust properties of differential rotation. These studies culminated in the consensus that solar internal rotation is zonal with a solidly rotating core, a differentially
rotating convection zone with two radial shear layers, 
one at the bottom of the convection zone,
termed the ``tachocline" \citep[][]{Spiegel-1992-A&A} and one near 
the surface, termed the near-surface shear layer
\citep[NSSL;][]{Thompson-1996-Sci}.
These radial shear layers have drawn attention
given their potential importance for driving the solar dynamo and their prominent role in maintaining the global angular momentum budget.

Quasi-degenerate perturbation theory \citep[QDPT;][, henceforth LR92]{lavely92}, on the other hand, accounts for cross-coupling between modes when computing frequency splittings. Although it represents a more accurate model, very few studies have employed QDPT in inferring DR because of its computational complexity as compared to DPT. \cite{schad20} formulated measurements in terms of mode-amplitude ratios to infer DR. \cite{Woodard-2013-SoPh} and \cite{Kashyap21} fit mode-amplitude spectra using the prescription of \cite{vorontsov11} (hereafter V11) to infer $a$-coefficients, which are polynomial expansion coefficients of frequency splittings. Higher-angular-degree modes are closely spaced in frequency, implying the worsening of the underlying assumption of DPT. \cite{Kashyap21} showed that the difference in splittings estimated by DPT and QDPT are statistically significant for larger angular degrees in the $f$ and $p_1$ branches. Here, we attempt to infer the rotation
profile through the application of the more general QDPT formalism, while still using the $a$-coefficients as our primary measurement. Given that only
high-angular degrees are coupled, we propose a ``hybrid" inversion method, where DPT is used for low-angular-degree modes and QDPT is used for high-angular-degree modes. Low-angular-degree modes are sensitive to greater depths, while the high angular degrees are trapped very close to the surface. Consequently, we expect only near-surface corrections from our study.

In this study, we use almost 22-years of $a$-coefficient measurements to infer DR. The period between May 1996 and April 2010 was constrained by measurements from MDI \citep{scherrer95}, onboard SOHO, and the period between April 2010 and January 2018 was constrained by measurements from HMI \citep{hmi}, onboard the SDO. While the odd $a$-coefficients are almost completely governed by differential rotation, the even $a$-coefficients contain contributions from structure perturbations such as solar oblateness \citep{woodard16}, magnetic fields \citep{Antia2000, Baldner2009}, sound-speed asphericity \citep{antia01, baldner2008} and second-order effects of rotation \citep{goughmag}. We carry out forward calculations via an exact eigenvalue solver to estimate first-order contributions to even $a$-coefficients due to solar rotation and compare our estimates with V11, which adopted a semi-analytic approach.

The outline of this paper is as follows. In Section~\ref{sec:theory},
we introduce basic notations for mode coupling in the context of DR as well as the $a$-coefficient formalism. Section~\ref{sec:inversion} furnishes details pertaining to (a) categorization of QDPT and DPT modes in Section~\ref{sec: mode_selection}, (b) setting up the cost-function to be minimized during inversion in Section~\ref{sec: inversion_methodology}, and, (c) determining optimal truncation in the angular degree of perturbation and the spectral window of coupled modes in Section~\ref{sec: smax_and_dlmax}. We present the results of corrections for time variation in DR in Section~\ref{sec: DPT_vs_hyb} and compare even-splitting coefficients with V11 results in Section~\ref{sec: even_acoeffs}. We summarize the key points in Section~\ref{sec:discussion}.
\newpage 

\section{Theoretical formulation: Isolated and coupled multiplets}
\label{sec:theory}

Linear perturbation analysis of the hydrodynamic equations of mass continuity,
conservation of momentum, and energy results in an eigenvalue problem which enables
calculation of eigenfrequencies and eigenfunctions of standard solar
models, such as Model-S \citep{jcd}. Such standard models do not account for effects of
asphericity, flows, anisotropy, non-adiabaticity, and magnetic fields. In the absence of such perturbations, these standard models have theoretically predicted ``degenerate" eigenfrequencies ${}_{n}\omega_{\ell}$ and eigenfunctions ${}_{n}\xi_{\ell m}$. Here $n$ is the radial order
and $\ell$ is the spherical harmonic degree. For a given multiplet $\mode{n}{\ell}$, the constituent $2\ell + 1$ modes
are labelled by a third quantum number $m\in [-\ell, \ell]$. The eigenfrequencies are degenerate in $m$ because of spherical symmetry.

The influence of the above-mentioned effects needs to be accounted for as additional
perturbations to the standard models. This results in splitting of eigenfrequencies and distortion of the eigenfunctions as follows
\begin{equation}
    {}_{n}\omega_{\ell} \to \omega_{\mathrm{ref}} + \delta {}_n\omega_{\ell m} \, , \qquad {}_{n}\bfxi_{\ell m} \to {}_{n}\bfxi_{\ell m} + \delta {}_{n}\bfxi_{\ell m} \, ,
\end{equation}
where $\delta {}_{n}\omega_{\ell m}$ and $\delta {}_{n}\bfxi_{\ell m}$ are the corrections to eigenfrequencies and eigenfunctions and $\omega_{\mathrm{ref}}$ is a reference frequency which maybe chosen to be close to the unperturbed frequency ${}_{n}\omega_{\ell}$ of the multiplet $\mode{n}{\ell}$ whose perturbed eigenstate we are interested in. This results a new eigenvalue problem for the ``supermatrix" $\sfZ$ (see Appendix~B in \cite{sbdas20})
\begin{equation} \label{eqn: eigenvalue_problem}
    \sum_{k \in \mathcal{K}} Z_{k'k} \, c_k = 2 \omega_{\mathrm{ref}} \, \delta \omega_{k'} \, c_{k'} \, .
\end{equation}
Elements of $\sfZ$ encode the coupling of modes in the presence of the aforementioned perturbations. Here, $k$ is conveniently used as a combined index to denote 
the mode $(n, \ell, m)$. For a temporal bandwidth $\Delta \omega$ and spectral bandwidth $\Delta \ell$, the supermatrix $\sfZ$ is built
around a ``central multiplet" $\mode{n_0}{\ell_0}$ by considering a set of modes $\mathcal{K}$ that obey $k \in \mathcal{K}$, $|{}_{n}\omega_{\ell} - {}_{n_0}\omega_{\ell_0}| < \Delta \omega$ and $|\ell - \ell_0| < \Delta \ell$. 
A sufficiently large $\Delta\omega$ and $\Delta\ell$
ensures that all significant couplings with $\mode{n_0}{\ell_0}$
are accounted for (see Section~\ref{sec: smax_and_dlmax}).
The perturbed eigenfrequencies of the $2\ell_0 + 1$ modes estimated from $\sfZ$ thereby accurately account for cross-coupling of multiplet $k_0$ with its proximal neighbours.
The unperturbed mass density is denoted by $\rho_0$ (Model S). Without loss of generality, we have chosen $\omega_{\mathrm{ref}} = \omega_{k_0}$, for our inversions.

Following the convention in LR92 and V11, we represent the 3D rotational velocity as
\begin{equation} \label{eq: vrot_VSH}
    \mathbf{v}_{\mathrm{rot}}(r,\theta,\phi) = - \sum_{s} w_s(r) \, \hat{\mathbf{r}} \times \bfnabla_1 Y_{s,0}(\theta,\phi) \, ,
\end{equation}
where $w_s(r)$ are the respective odd-degree toroidal coefficients and $Y_{\ell, m}(\theta, \phi)$ are spherical harmonics labelled by angular degree $\ell$ and azimuthal order $m$. The elements of the supermatrix due to the rotation field $\bfv_{\mathrm{rot}}$ may be expressed as (see Eqns.~[135-136] in LR92 and Eqn.~[A1] in V11)
\begin{equation} \label{eqn:DR_coupmat_V11}
\begin{split}
    Z^{(n_0,\ell_0)}_{k'k,m} = 2\omega_{\mathrm{ref}} \, \kappa_{\ell'} \kappa_{\ell}
    \sum_{s} \gamma_s \, \wigfull{\ell'}{s}{\ell}{1}{0}{-1}
    \wigfull{\ell'}{s}{\ell}{m}{0}{-m} \int_{r=0}^{R_{\odot}} \rho_0 \,
    w_s(r) \, T^{k'k}_s(r) \, r \, \rmd r + \left(\omega_k^2 -
    \omega_{\rm{ref}}^2\right) \, \delta_{k'k}.
\end{split}
\end{equation}
Note that the superscript $(n_0,\ell_0)$ denotes the particular central multiplet for which the supermatrix is constructed and the subscript $m$ is used to explictly imply that only modes with the same azimuthal order are coupled in the presence of an axisymmetric flow field. We also have
$\kappa_{\ell} = \left[\ell (\ell+1) (2\ell+1) \right]^{1/2}$, $\gamma_s = \sqrt{(2s + 1)/4\pi}$ and
$T^{k'k}_s(r)$, the sensitivity kernel for $w_s(r)$, given by
\begin{equation}
    T^{k'k}_{s}(r) = \left[U_{k'} U_{k} - U_{k'}V_k - V_{k'}U_k
    +\frac{\ell'(\ell'+1) +
    \ell(\ell+1) - s(s+1)}{2} V_{k'}V_k \right].
\end{equation}
$U_k(r)$ and $V_k(r)$  are the radial
and horizontal eigenfunctions corresponding to mode $k$, respectively.
$T_s^{k'k}(r_0)$ therefore encodes the degree to which the multiplets $k$
and $k'$ couple in the presence of an axisymmetric, degree-$s$ rotation field of unit strength at $r=r_0$.

\begin{figure}
    \centering
    \includegraphics[width=0.5\textwidth]{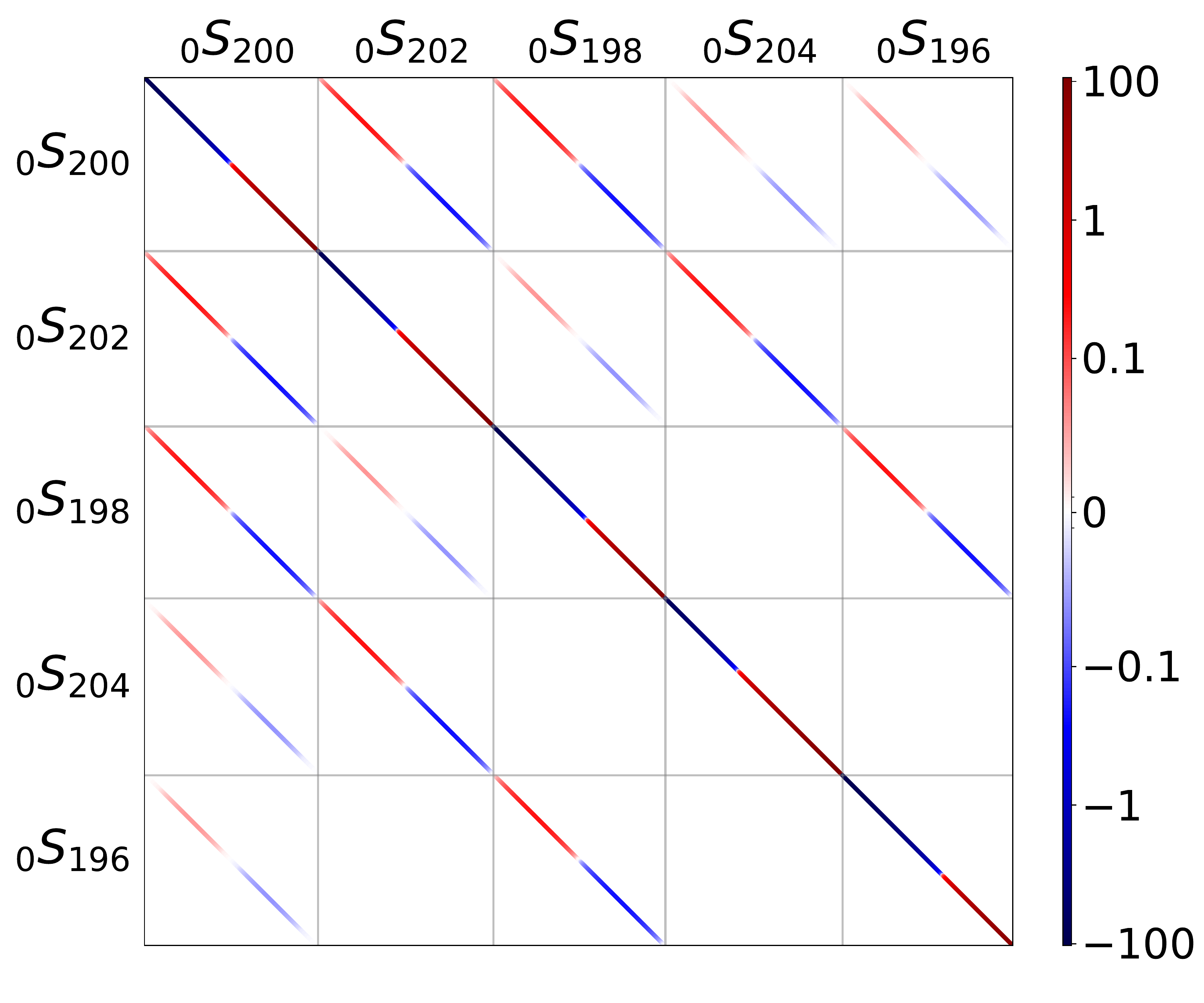}
    \caption{Supermatrix corresponding to the central multiplet $\mode{0}{200}$. The colors are spaced out in logarithmic scale for better visibility of the weak off-diagonal coupling components. Grids in the matrix denote the submatrices capturing coupling of multiplets, which can be read off from the \textit{top} of the column and \textit{left} of the row. $\mode{0}{200} - \mode{0}{|200 \pm 2|}$ couplings are 100~times weaker, whereas $\mode{0}{200} - \mode{0}{|200 \pm 4|}$ couplings are 1000~times weaker than self-coupling. We have used the \texttt{symlog} functionality of the \texttt{matplotlib} Python package to represent negative values in logarithmic scale.}
    \label{fig:supmat}
\end{figure}

The eigenvalues of acoustic modes in Model-S are degenerate in $m$, i.e., ${}_{n_0}\omega_{\ell_0 m} = {}_{n_0}\omega_{\ell_0}.$
Differential rotation is an axisymmetric perturbation that breaks the
spherical symmetry of the system, thus lifting the degeneracy in $m$ and ``splitting" the frequencies. As shown in Eqn.~(\ref{eqn: eigenvalue_problem}), the frequency splitting associated with mode $k_0$ may be estimated from
eigenvalues of the supermatrix $Z^{(n_0, \ell_0)}_{kk', m}$. The supermatrix for $\mode{0}{200}$ due to differential rotation, considering couplings for $196 \leq \ell \leq 204$, is shown in Fig.~\ref{fig:supmat}. 

It is standard practice in helioseismology to project frequency 
splittings $\delta{}_n\omega{}_{\ell m} = {}_n\omega{}_{\ell m}-{}_n\omega{}_{\ell}$ on to a basis of orthogonal polynomials $\cP_j(m)$. The resultant fitting coefficients $a_j^{n\ell}$ are the so-called
``$a$-coefficients":
\begin{equation}\label{eqn:freq_splitting}
{}_n\omega{}_{\ell m} = {}_n\omega{}_{\ell} + \sum_{j=0}^{j_\text{max}} a^{n\ell}_j\, \cP_j(m).
\end{equation}
In practice, $a$-coefficients
are recorded for $j_\text{max}=36$ \citep[e.g.,][]{schou_data}. A recipe for
obtaining these may be found in Appendix~A of \cite{schou_pol_94}. 
Harnessing the orthonormality of the basis polynomials, $\sum_{m=-\ell}^\ell
\cP_{j}(m) \cP_{k}(m) = \delta_{jk}$, we write the $a$-coefficients as
\begin{equation}\label{eqn:a_def}
a^{n\ell}_j = \sum_{m=-\ell}^\ell \delta{}_n\omega{}_{\ell m}\,\cP_j(m).
\end{equation}

The isolated multiplet approximation for a zonal perturbation implies that
the frequency splittings $\delta_n \omega_{\ell m}$ in Eqn.~(\ref{eqn:a_def}) are 
equal to the diagonal elements of $Z^{(n_0,\ell_0)}_{k_0 k_0, m}$ scaled by
$1/(2 \, {}_{n_0}\omega_{\ell_0})$. While this is a good approximation for low 
angular-degree multiplets, \cite{Kashyap21} showed that there is an 
error (in an L2-norm sense) that is significant enough to be detectable
(about 2$\sigma$), when using the 
isolated-multiplet approximation for high-$\ell$ multiplets. 
This error arises when the off-diagonal components of the supermatrix 
(as shown in Fig.~\ref{fig:supmat} for the central multiplet $\mode{0}{200}$) are ignored. The rest of the paper will explore the effects of considering these off-diagonal elements
on the $a$-coefficients predicted from forward calculations and the systematic changes in 
inverted profiles from 22 years of MDI and HMI measurements.

\section{Inversion} \label{sec:inversion}
\subsection{Selection of coupled multiplets} \label{sec: mode_selection}

We carry out inversions for differential rotation using $a$-coefficients obtained from 72-day data-sets 
over the period between 1996-05-01 and 2018-01-06 (all dates in this study
use the \texttt{yyyy-mm-dd} format). We do away with the isolated multiplet
approximation for modes with $\ell\geq160$ in the $n=0, 1$ radial branches.
The L2-norm error decreases rapidly with decreasing
$\ell$ \citep{Kashyap21}. High-precision observations from HMI
possess uncertainties as small as 0.03\% of the observed 
$a$-coefficients. Hence, a safe cutoff for the L2-norm error is 
taken to be $10^{-4}$ from Fig.~9 of \cite{Kashyap21}, and all modes
with larger errors are considered for the full-coupling problem.
Henceforth, we refer to these as ``coupled" multiplets, while the others
are ``isolated" multiplets. The inversions are
``hybrid" in nature since frequency splittings are modeled in two ways: 
\begin{enumerate}[(A)]
    \item For isolated multiplets, the main diagonal of the submatrix 
    corresponding to the central multiplet $k_0$ is used, i.e.,
    $\delta_n \omega_{\ell m} = $ \texttt{diag}$\left(Z_{k_0,k_0}^{(n_0,\ell_0)}\right)/(2 \, {}_{n_0}\omega_{\ell_0})$.
    We drop the superscript label $(n_0,\ell_0)$ on the supermatrix from now on. It is 
    implied that any supermatrix corresponds to a central multiplet $\mode{n_0}{\ell_0}$.
    \item For coupled multiplets, the eigenvalue problem is solved 
    for the entire supermatrix $\sfZ$ and frequency splittings 
    corresponding to the central multiplet $\mode{n_0}{\ell_0}$ are read off.
    This one-to-one mapping between eigenfrequencies and the corresponding 
    modes is possible since the perturbed eigenfunctions are very close 
    to the unperturbed analogs --- arising from the diagonally dominant
    nature of the supermatrix. We use \texttt{float32} for all computation,
    since any difference with higher precision is significantly lower
    than observational noise.
\end{enumerate}
In hybrid inversions, based on our defined cutoff, we see that the typical
number of observed multiplets is $\sim 1800$, and $\sim 200$ of those 
are found to be coupled, i.e., roughly 10\% of all the observed modes 
are treated as coupled, and these are
responsible for corrections to the DPT-rotation profiles. 
To understand the depth up to which coupled multiplets can result
in such corrections, we first estimate the lower turning points
of coupled modes. Invoking the Cowling approximation, it may be shown 
that the lower turning point  $r_{\mathrm{LTP}}$ of modes ($n>0$) may be 
expressed as 
\begin{equation}\label{eq: rltp}
    \frac{c^2(r_{\mathrm{LTP}})}{r_{\mathrm{LTP}}^2} = \frac{\omega^2}{\ell \, (\ell + 1)},
\end{equation}
where $c(r)$ is the sound speed at radial distance $r$ from the center of the
Sun and $\omega$ the mode frequency. The
lowest angular degree for the two radial branches $n=0,1$ where mode-coupling is considered is $\ell = 160$. The above equation is valid for $n>0$ radial orders. So, using Eqn.~(\ref{eq: rltp}) to estimate the $r_{\mathrm{LTP}}$ for $\mode{1}{160}$, we get $r^{n=1}_{\mathrm{LTP}}=0.971\,R_\odot$. Since eigenfunctions corresponding to the $n=0$ branch, has no nodes in radius, it peaks close to the surface and dies off in an evanescent manner. At $r = 0.971\,R_\odot$, the $\mode{0}{160}$ mode eigenfunctions are approximately two orders of magnitude weaker than the maxima near the surface. Going deeper to $r = 0.9\,R_\odot$, we see that these eigenfunctions are six orders of magnitude smaller than the near-surface maximum. It is
therefore expected that differential rotation below $0.9\,R_\odot$ is
weakly sensitive to the effect of coupled modes. In our inversions, we fit for
differential rotation in the spherical shell bounded by $r \in 
[0.9\,R_{\odot}, 1.0\,R_\odot]$.

\subsection{Inversion methodology} \label{sec: inversion_methodology}
We parameterize $w_s(r)$ on a radial grid on a basis comprising cubic B-splines $\beta_p(r)$, where index $p$ refers to the knot location corresponding to a specific B-spline polynomial with coefficient $c_s^p$, i.e.,
\begin{equation} \label{eqn:spline_decomposition}
    w_s(r) = 
    \begin{cases}
    \sum_p c_s^{p, \, \mathrm{fixed}} \, 
    \beta_p(r) \qquad \mathrm{for} \qquad r < 0.9 \\
    \sum_p c_s^p \, \beta_p(r) \qquad \qquad \mathrm{for} \qquad r \ge 0.9
    \end{cases},
\end{equation}
where $c_s^{p, \, \mathrm{fixed}}$ cooresponds to the spline
coefficients of the profiles of 2D RLS inversions.
We use the same radial grid as the 2D RLS inversions hosted on 
Stanford University's Joint Science Operations Center (JSOC) database
\footnote{\href{http://jsoc.stanford.edu/SUM95/D892366801/S00000/rmesh.orig}{Radial grid on JSOC.}}.
Therefore, the inverse problem of finding $\Omega(r,\theta,\phi)$ reduces to inferring the coefficients $c_s^p$. Using Eqns.~(\ref{eqn:DR_coupmat_V11}) and (\ref{eqn:spline_decomposition}), we  express the supermatrix more explicitly in terms of the spline coefficients as
\begin{equation} \label{eqn:supmat_in_c}
    Z_{k'k} = \sum_p \sum_{s=1,3,5} c_s^p\, \Lambda_{p, k'k}^s + \left(\omega_k^2 -
    \omega_{\rm{ref}}^2\right) \, \delta_{k'k},
\end{equation}
where,
\begin{equation}
    \Lambda_{p, k'k}^s = 2\,\omega_{\mathrm{ref}} \, \kappa_{\ell'}\, \kappa_{\ell} \,\gamma_s \, \wigfull{\ell'}{s}{\ell}{1}{0}{-1}
    \wigfull{\ell'}{s}{\ell}{m}{0}{-m} \int_{r=0}^{R_{\odot}} \rho_0(r) \,
    \beta_p(r) \, T^{k'k}_s(r) \, r\, \rmd r.
\end{equation}
The separation of terms that do not depend on $c_s^p$ into $\Lambda_{p,k'k}^s$ is essential. This allows for a one-time precomputation of $\Lambda_{p,k'k}^s$, 
saving both time and computational expense during the non-linear inversion when 
considering coupled modes. Since we only fit for differential rotation for $r 
\geq 0.9$, we classify the spline coefficients $c_s^p$ into a fixed component, 
denoted by $c_s^{p, \, \mathrm{fixed}}$, corresponding to the basis functions 
with local support in $r < 0.9$ and the inverted component, denoted by $c_s^{p, 
\, \mathrm{fit}}$, corresponding to the basis functions with local support in $r 
\geq 0.9$. The $c_s^{p, \, \mathrm{fixed}}$ is computed using profiles available on 
JSOC.
Accordingly, $\Lambda_{p, k'k}^s$ would also split up into 
$\Lambda_{p, k'k}^{s, \, \mathrm{fixed}}$ and $\Lambda_{p, k'k}^{s, \, 
\mathrm{fit}}$. The supermatrix in Eqn.~(\ref{eqn:supmat_in_c}) may then be 
expressed in terms of a fixed part $Z_{k'k}^{\mathrm{fixed}}$ and the $c_s^{p, \, \mathrm{fit}}$ dependent part,
\begin{equation}\label{eqn: build_full_supmat}
    Z_{k'k} = Z_{k'k}^{\mathrm{fixed}} + \sum_p \sum_{s=1,3,5} c_s^{p, \, \mathrm{fit}}\, \Lambda_{p, k'k}^{s, \, \mathrm{fit}},
\end{equation}
where $Z_{k'k}^{\mathrm{fixed}} = \sum_p \sum_{s=1,3,5} c_s^{p, \, \mathrm{fixed}}\, \Lambda_{p, k'k}^{s, \, \mathrm{fixed}} + \left(\omega_k^2 - \omega_{\rm{ref}}^2\right) \, \delta_{k'k}$ and is precomputed along with $\Lambda_{p, k'k}^{s, \, \mathrm{fit}}$.
Finally, the modeled data $d_{\mathrm{mod}}$ is expressed as 
\begin{equation} \label{eqn: model_DPT_hybrid}
d_{\mathrm{mod}} = \begin{cases}
\cA\left[\texttt{diag}(Z)\right]/(2\,\omega_{k_0}) &\text{for isolated modes,}\\
\cA\left[\texttt{eig}(Z)\right]/(2\,\omega_{k_0}) &\text{for coupled modes},
\end{cases}
\end{equation}
where $\texttt{diag}(Z)$ returns elements on the main diagonal of the supermatrix $Z$ corresponding to the self-coupling of the central multiplet $\mode{n_0}{\ell_0} - \, \mode{n_0}{\ell_0}$, and $\texttt{eig}(Z)$ returns eigenvalues corresponding to the central multiplet $\mode{n_0}{\ell_0}$ after solving the eigenvalue problem for the supermatrix $Z$. The operator $\mathcal{A}$ indicates the selection of modes belonging to the central multiplet $\mode{n_0}{\ell_0}$.

For the inversion, we define a regularized misfit function $\chi^2$ in an L2-norm sense
\begin{equation}
        \chi^2 = \overbrace{\sum_{i \, \in \, \mathrm{all\,modes}} \left(\frac{d^i - d^i_{\mathrm{mod}}}{\sigma_d^i}\right)^2}^{\chi^2_d = \text{data misfit}} + \sum_{s=1,3,5,7,9} \mu_s \overbrace{\left[ \frac{\rmd^2 w_s(r)}{\rmd r^2}\right]^2}^{\chi^2_m = \text{model smoothness}},
\end{equation}
where $d_i$ and $\sigma_d^i$ are the $a$-coefficients and their corresponding uncertainties, measured from HMI or MDI, and $\mu_s$ are angular-degree-dependent regularization parameters. The second term dictates the smoothness of the inverted profile. When using splines, it may be shown that this term can be recast in terms of the spline coefficients $c_s^p$ and an operator $D$ which captures the second derivative of the basis functions $\beta_p(r)$ with respect to $r$:
\begin{equation}
    \left[ \frac{\rmd^2 w_s(r)}{\rmd r^2}\right]^2 = 
    \sum_{i,j \in \mathrm{all\,knots}} c_s^i \, D^{ij} \, c_s^j.
\end{equation}
We further use a non-dimensional regularization parameter $\overline{\mu}_s$, which is related to $\mu_s$ as below
\begin{equation}
    \mu_s = \overline{\mu}_s\,N \, \frac{\mathrm{tr}(H_d)}{\mathrm{tr}(H_m)},
\end{equation}
where $N$ is the total number of data $a$-coefficients $d^i$,  $H_d$ the Hessian associated with the data misfit $\chi^2_d$, and $H_m$ the Hessian associated with the model misfit $\chi^2_m$. These Hessians are computed as second derivatives of the respective $\chi^2$ with respect to the vector of the fitted spline coefficients $c_s^{p, \, \mathrm{fit}}$. 
The inverse problem is non-linear due to the eigenvalue operation as shown in Eqn.~(\ref{eqn: model_DPT_hybrid}) and the model parameters $c_s^{p, \, \mathrm{fit}}$ undergo an iterative march towards the final solution via the standard Newton's method to minimize the misfit, $\chi^2$ \citep{Tarantola-1987-iptm}. The marching in parameter-space requires the computation of gradient and hessian. These quantities are computed numerically using the \texttt{autograd} functionality of \texttt{jax} \citep{jax2018github}. Further discussions on the accuracy and robustness on the eigenvalue operator and inversion tools can be found in Appendix~\ref{sec: eig_accuracy}.

The non-dimensional regularization parameter $\overline{\mu}_s$ used for the inversions presented in this paper is chosen by the traditional L-curve method. For this, we chose 50 logarithmically-spaced values of $\overline{\mu}_s$ in a sufficiently large window of $[10^{-6}, 10^6]$ for each $s$. For each of these values of $\overline{\mu}_s$, we perform the inversion and calculate the data misfit $\chi_d^2$ and model misfit $\chi^2_m$ from the inverted model parameters. Subsequently, we made the L-curve by plotting $\chi^2_d$ vs. $\chi^2_m$ and interpolated the curve with smooth cubic splines using $\texttt{interp1d}$ function of Python's $\texttt{scipy}$ module. The $\overline{\mu}_s$ for the final inversion results presented in this paper was chosen from the knee of this L-curve. A representative plot for the L-curve and a demonstration of the convergence for our hybrid inversions is provided in Appendix~\ref{sec: inversion_validation}.

\subsection{Determining $s_{\mathrm{max}}$ of $w_s(r)$ and supermatrix dimensionality} \label{sec: smax_and_dlmax}

\begin{figure}
    \centering
    \includegraphics[width=\textwidth]{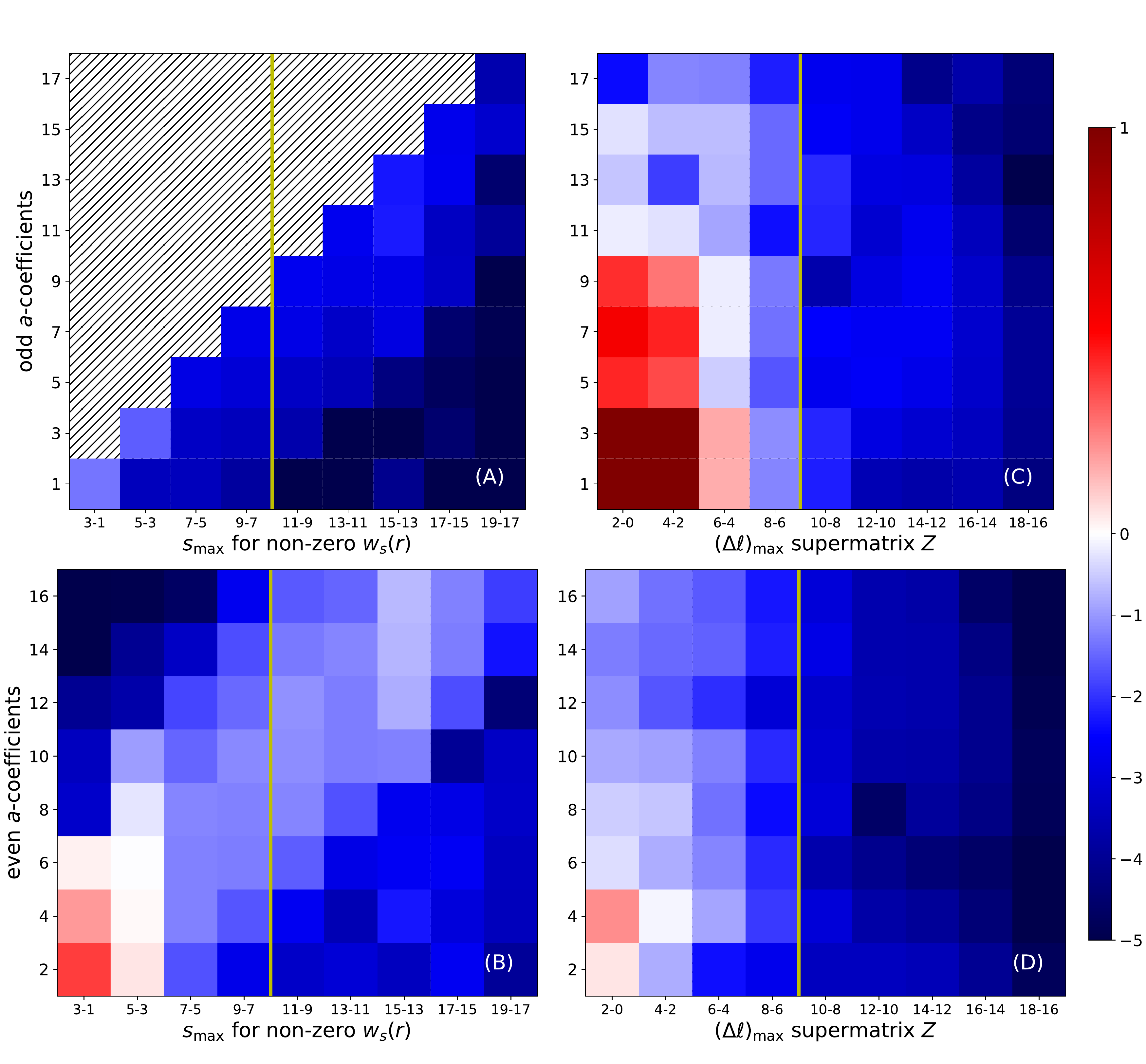}
    \caption{\textit{Left} panel: Consecutive differences of $a$-coefficients when considering increasing $s_{\mathrm{max}}$ for non-zero $w_s$. To illustrate further, in (B), the \textit{red} block on the lower left shows the quantity $\log_{10}\left\{[a_2(s_{\mathrm{max}}=3) - a_2(s_{\mathrm{max}}=1)]/\sigma(a_2)\right\}$. These consecutive differences for the odd and even $a$-coefficients are shown in the \textit{top} and \textit{bottom} rows, respectively. All blocks in the area to the right of the \textit{yellow} line demarcate statistically insignificant consecutive differences. \textit{Right} panel: Same as the \textit{left} panel, but for an increasing number of neighbouring multiplets $\Delta \ell_{\mathrm{max}}$ instead of increasing maximum-angular-degree $s_{\mathrm{max}}$. Forward calculations were done for the multiplet $\mode{0}{280}$ using inverted rotation profiles available on JSOC, which uses HMI data measured during the 72-day solar minima period between 2019-03-14 and 2019-05-25. Since this figure demonstrates the saturation of $a$-coefficients and odd $a$-coefficients $a_j$ are significant only when $j \geq s_{\mathrm{max}}$, we \textit{hatch} the upper-triangle in (A).}
    \label{fig:2D_smax_jmax}
\end{figure}

Coupling of modes is permissible within a certain frequency window $\Delta \omega$. Further, the Wigner $3-j$~symbols in Eqn.~(\ref{eqn:DR_coupmat_V11}) impose the selection rule $|\ell - \ell_0| \leq s$, which disallows a central multiplet with angular degree $\ell_0$ from coupling with modes outside the window $\ell' \in [\ell_0 - s, \, \ell_0 + s]$. This means that submatrices corresponding to coupling of angular degrees $(\ell_0, \ell')$ would be non-zero. Consequently, with an increase in the maximum angular degree of rotation $s_{\mathrm{max}}$, farther and farther bands in the supermatrix $\sfZ$ are filled. The total number of submatrices that constitutes a supermatrix, depends on the angular degree of the multiplet farthest from the central multiplet. We define this maximum offset in angular degree between the central multiplet and its farthest neighbour as $\Delta \ell_{\mathrm{max}} = \mathrm{max}(\ell - \ell_0$). Although at first it might seem that $s_{\mathrm{max}} = \Delta \ell_{\mathrm{max}}$, that is not true. $\Delta \ell_{\mathrm{max}}$, infact, controls the frequency window $\Delta \omega$ and in order for the eigenfrequencies to converge to a stable value, it is essential to choose a large enough $\Delta \omega$ and hence a large enough $\Delta \ell_{\mathrm{max}}$. The supermatrix would therefore have non-zero bands upto the submatrices where $\ell \leq \ell_0 \pm s$ and have zeros in all submatrices thereafter. Since each multiplet contains $2\ell + 1$ modes, large values of $s$ or $\Delta \ell_{\mathrm{max}}$ correspond to large sizes of supermatrix $\sfZ$. The non-linear inversion involves solving these eigenvalue problems for all coupled multiplets at every iteration. Consequently, with increasing size of $\sfZ$, the computational cost of inversions grows significantly. In principle, choosing $s_{\mathrm{max}}$ and $\Delta \ell_{\mathrm{max}}$ ensures all components of differential rotation has been accounted for as well as coupling with all modes have been considered. Therefore, the ``true" estimate of the perturbed eigenstates require $s_{\mathrm{max}}, \, \Delta \ell_{\mathrm{max}} \to \infty$. However, this is (A) not computationally tractable, and (B) not practically necessary since the eigenfrequencies and eigenfunctions converge to a stable solution with increasing values of both $s_{\mathrm{max}}$ and $\Delta \ell_{\mathrm{max}}$. This necessitates a judicious choice of truncation of the maximum angular degree $s_{\mathrm{max}}$ of $w_s(r)$ and the farthest neighbours in $\ell$ that need to be considered for the eigenvalue problem to converge sufficiently. Since the maximum coupling is expected where $(\partial \omega/\partial \ell)_n$ is the smallest along a radial branch $n$, we choose $\ell_0 = 281$ on the $f$-branch \citep[see Fig.~1 in][]{gizon05}. Since we use modes upto $\ell=300$ on the radial branch with $\Delta \ell_{\mathrm{max}}=19$, $\mode{0}{281}$ is the maximally-coupled central multiplet that could be analyzed.

We test the optimal $s_{\mathrm{max}}$ for differential rotation via forward calculations using 2D-RLS profiles from JSOC during the solar minimum corresponding to the 72-day period between 2019-03-14 and 2019-05-25. To do this, we first vary $s_{\mathrm{max}}$ from 1 to 19 holding $\Delta \ell_{\mathrm{max}} = 19$ fixed. Although modes beyond $\ell_0 \pm s_{\mathrm{max}}$ do not couple with the central multiplet, this test allows us to check for convergence in the $a$-coefficients as a function of $s_{\mathrm{max}}$. The result of this test is summarized in the \textit{left} panel of Fig.~\ref{fig:2D_smax_jmax}. Each tile in the colour-map represents the quantity $\log_{10}\left\{[a_j(s_{\mathrm{max}}=n+2) - a_j(s_{\mathrm{max}}=n)]/\sigma(a_j)\right\}$. This serves as a measure of the amount of change in the $a$-coefficients when $s_\mathrm{max}$ is increased by 2. The difference is scaled by the observed uncertainty of the $a$-coefficients to indicate whether the changes are statistically significant. We have used a diverging colourbar with \textit{red} patches indicating non-negligible differences between successive $s_{\mathrm{max}}$ cases, while \textit{blue} patches indicate negligible differences. Fig.~\ref{fig:2D_smax_jmax}(A) shows that, for odd $a$-coefficients, the consecutive differences are negligible for all $s_{\mathrm{max}}$. For completeness, we also present the consecutive differences of the even $a$-coefficients in Fig.~\ref{fig:2D_smax_jmax}(B). It may be safely inferred that from the case ``11-9" onwards, the consecutive differences are negligible. Therefore, considering $w_s(r) = 0$ for $s > 9$ should give us $a$-coefficients that are not statistically different from solving the problem with $s_{\mathrm{max}}=19$.

Next, we vary $\Delta \ell_{\mathrm{max}}$ from 1 to 19 holding $s_{\mathrm{max}} = 19$ fixed. This serves as an independent test for how many neighbouring multiplets we need to consider in order to ensure saturation of $a$-coefficients. The results are summarized in the \textit{right} panel of Fig.~\ref{fig:2D_smax_jmax}. Similar to the previous test, the coloured tiles represent the quantity $\log_{10}\left\{[a_j(\Delta \ell_{\mathrm{max}}=n+2) - a_j(\Delta \ell_{\mathrm{max}}=n)]/\sigma(a_j)\right\}$. From Fig.~\ref{fig:2D_smax_jmax}(C) \& (D), we see that all red patches occur for $\Delta \ell_{\mathrm{max}}\leq8$.

\section{Results}
\label{sec:results}

\subsection{Inverse problem: DPT vs. hybrid} \label{sec: DPT_vs_hyb}
\begin{figure}
    \centering
    \includegraphics[width=\textwidth]{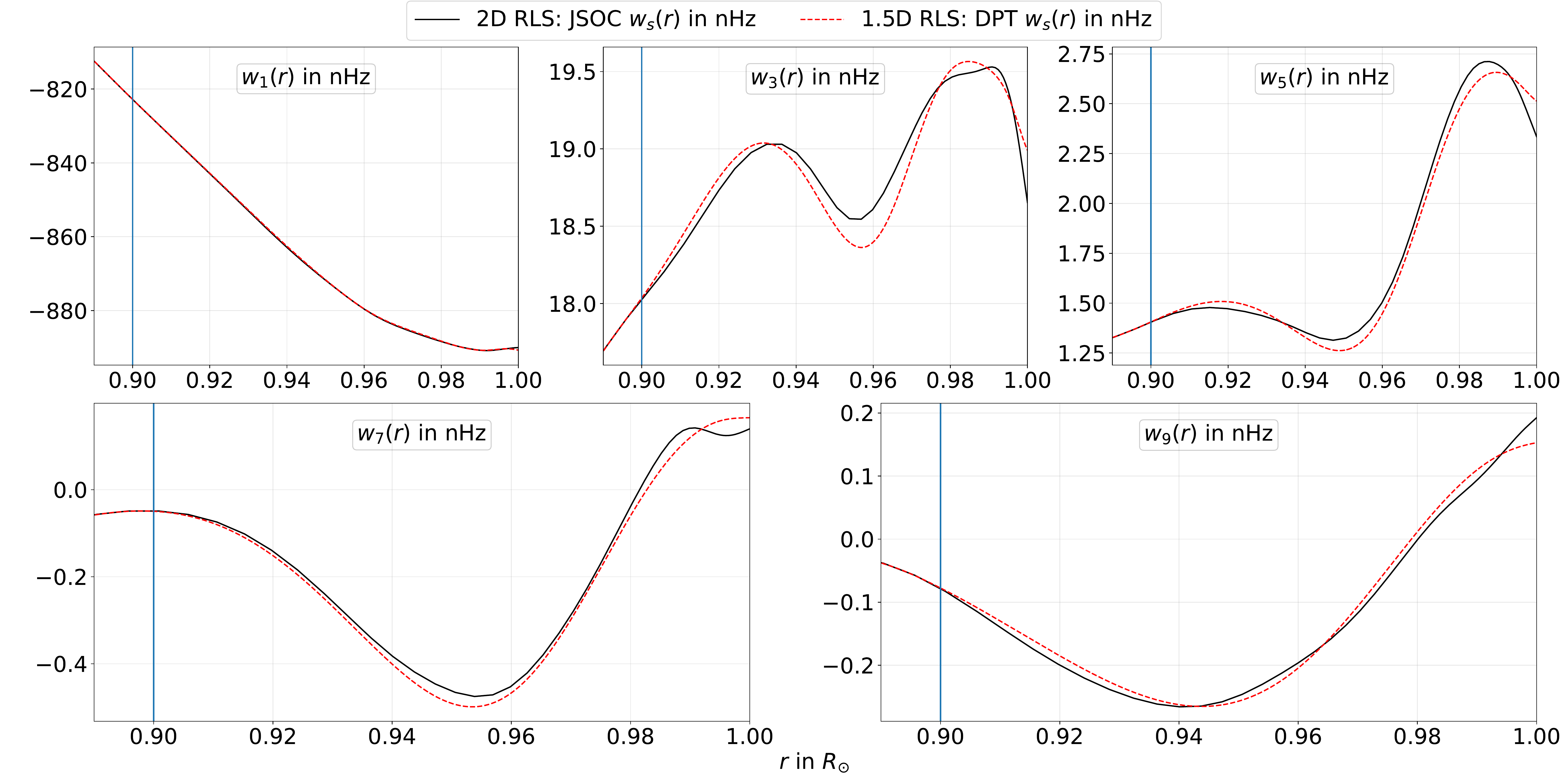}
    \caption{Comparison between differential-rotation profiles $w_s(r)$ up to angular degree $s=9$ obtained from JSOC in \textit{black} and our DPT inversions in \textit{dashed red} lines. The JSOC profiles are labelled ``2D RLS" since they were obtained from projecting the 2D inversions onto a 1.5D basis, as discussed in Appendix~\ref{sec: JSOC_to_1pt5D}. Since the modes that we use in hybrid inversions are almost insensitive below $r_{\mathrm{th}}=0.9 R_{\odot}$, we fix $w_s(r)$ to the JSOC values below this depth, as indicated by the vertical \textit{blue} lines.}
    \label{fig: JSOC_vs_DPT}
\end{figure}

\begin{figure}[h]
\centering
\includegraphics[width=\textwidth]{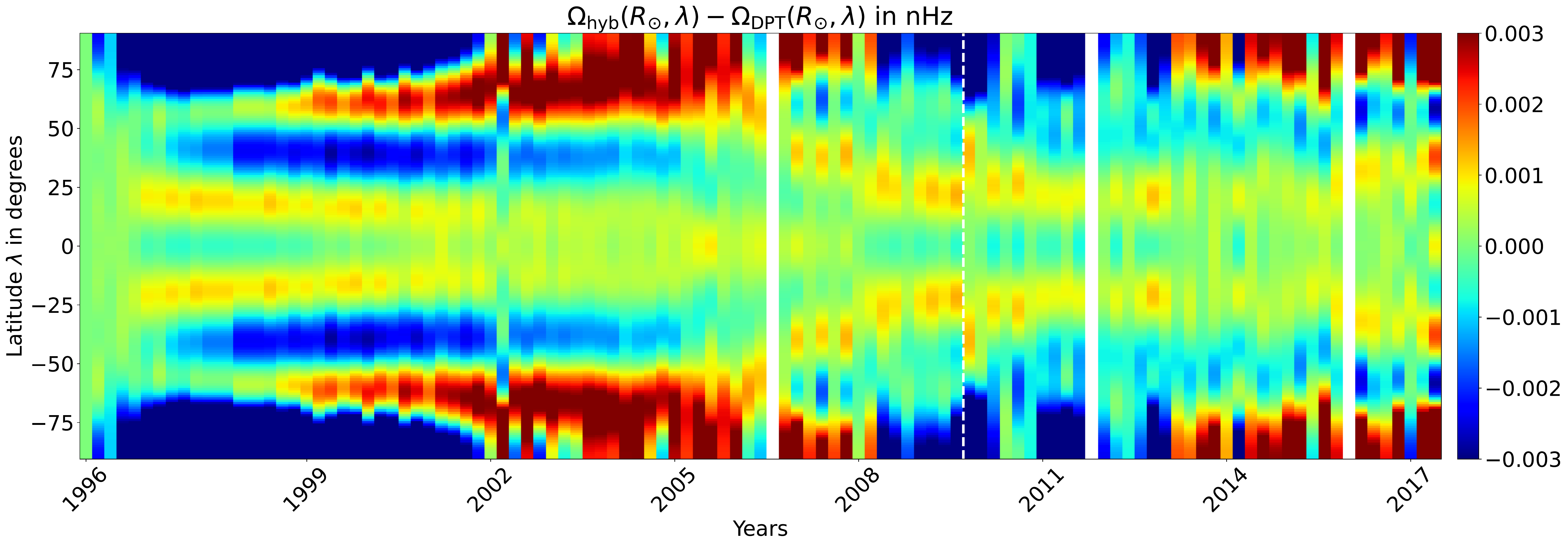}
\caption{Offset in surface differential rotation inferred using hybrid inversions as compared to DPT inversions spanning 22 years. We have used MDI measurements from May 1996 to April 2010 and thereafter HMI measurements up to January 2018. The vertical \textit{dashed-white} line indicates the separation between MDI and HMI regions.}
\label{fig: timeseries_lat}
\bigbreak
\includegraphics[width=\textwidth]{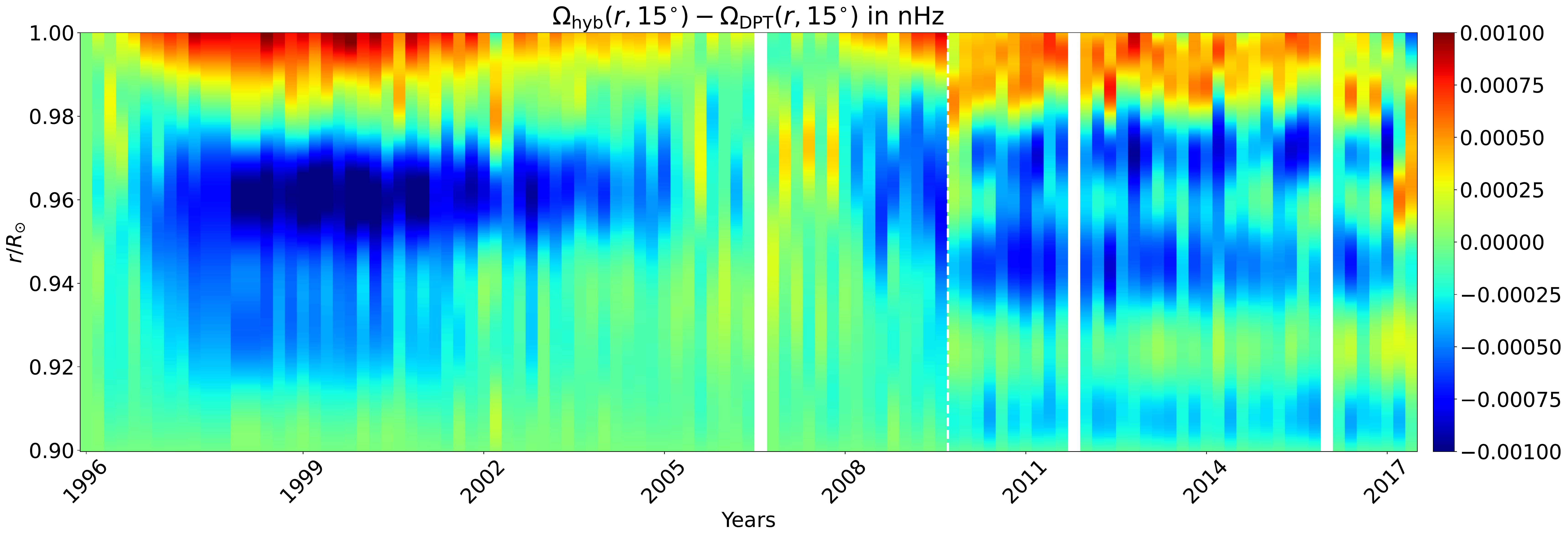}
\caption{Offset in 22 years of differential rotation as a function of depth for the chosen latitude $15^{\circ}$. As in Fig.~\ref{fig: timeseries_lat}, the vertical \textit{dashed-white} line marks the separation between MDI and HMI regions.}
\label{fig: timeseries_depth}
\end{figure}

As demonstrated in Section.~\ref{sec: smax_and_dlmax}, we may choose $s_{\mathrm{max}} = 9$ and $\Delta \ell_{\mathrm{max}} = 8$ without incurring any statistically significant errors due to truncation. The resultant supermatrix $\sfZ^{n_0, \ell_0}$ contains contributions from perturbations due to $w_1(r)$ through $w_9(r)$ which couples neighbouring multiplets in the spectral window $\ell_0 - 9 < \ell < \ell_0 + 9$. As illustrated in Eqn.~(\ref{eqn: model_DPT_hybrid}), we model the data as the diagonal elements of $\sfZ$ for the isolated modes. For coupled modes, the data is modeled via the full eigenvalue solver. As is customary, we have validated our inversion methodology with synthetic data (with and without noise) in Appendix~\ref{sec: inversion_validation}. Having presented this verification of our hybrid inversion methodology using artificially generated data for a known rotation profile, we proceed to perform hybrid inversions on real data.

We carry out the inversion using this ``hybrid" modeling method and compare the results against a purely DPT inversion, which is linear and considers all multiplets as isolated. The former represents an exact model and consequently an ``exact" inversion, while the latter is an approximation that has been traditionally used in inferring differential-rotation profiles.

In Fig.~\ref{fig: JSOC_vs_DPT}, we show the consistency of our 1.5D DPT inversions as compared to the 2D RLS inversions available on JSOC, both of which ignore cross-coupling of modes. The \textit{black} line is a projection of the rotation profile $\Omega(r,\theta)$ inferred via 2D RLS onto the 1.5D profile $w_s(r)$, as outlined in Appendix~(\ref{sec: JSOC_to_1pt5D}). The \textit{red} line is from our 1.5D inversions. We note that the two profiles are not expected to match exactly because of differences such as inversion methodologies, model parameterization and exact choice of regularization parameters. Despite this, the two profiles agree well both qualitatively and quantitatively. The plotted $w_s(r)$ profiles are from the 72-day MDI $a$-coefficient measurement starting on $12^{\mathrm{th}}$ July, 1996.

Having established the validity of our DPT inversions against traditionally accepted results, we present the differences between differential rotation profiles inferred from our hybrid and DPT inversions in Figs.~\ref{fig: timeseries_lat} and \ref{fig: timeseries_depth}. Fig.~\ref{fig: timeseries_lat} shows the difference at the solar surface as a function of latitude over 22 years, in chunks of 72-day periods, between May 1996 and January 2018. For this, we ignore temporal variations of differential rotation within each 72-day period. We perform separate inversions of $\mathbf{v}_{\mathrm{rot}}$ (see Eqn.~[\ref{eq: vrot_VSH}]) from $a$-coefficient measurements estimated from the corresponding 72-day time-series. Fig.~\ref{fig: timeseries_depth} is the same, but plotted as a function of depth from the solar surface down to $0.9 R_{\odot}$ at latitude $\lambda = 15^{\circ}$. We used MDI measurements up to April 2010 and HMI measurements thereafter. This is indicated by the vertical \textit{dashed-white} line in both figures.  Comparing our Figs.~\ref{fig: timeseries_lat} and \ref{fig: timeseries_depth} with Figs.~3(C) and~3(D) in \cite{vorontsov02}, respectively, we note that the correction in differential rotation due to mode coupling is approximately three orders of magnitude smaller than the torsional-oscillation signal.

Despite this small difference, we note a systematic pattern as a function of time. In Fig.~\ref{fig: timeseries_lat}, for the MDI years, we see bands of fast rotation around $20^{\circ}$ latitudes and slow rotation around $40^{\circ}$ latitudes. The poles show an alternating pattern of 5~years of slow followed by fast rotation. Similar features can be discerned in the HMI years. However, the bands of slower rotation, which were at $40^{\circ}$ for MDI, are less pronounced and seem to have moved to around $60^{\circ}$ for HMI. Moreover, the sub-polar branches of fast rotation in MDI that start around 1998 and merge with the polar fast-rotation band around 2002 are missing for the HMI years (which were expected between 2011 and 2014). This discrepancy could possibly be attributed to the change in instruments and the lower SNR for $a_7^{n \ell}$ in HMI as compared to MDI measurements. The depth profile in Fig.~\ref{fig: timeseries_depth} broadly shows a layer of faster rotation around $0.98 R_{\odot} < r < 1.0 R_{\odot}$ and a layer of slower rotation between $0.93 R_{\odot} < r < 0.97 R_{\odot}$. Once again, there are differences between the MDI and HMI years, which may be (at least in part) attributed to systematic differences in measurements. Although these demonstrate that there are systematic differences that may be spotted upon careful scrutiny, these are minuscule and may be disregarded for practical purposes.

\subsection{Effect on even $a$-coefficients} \label{sec: even_acoeffs}

\begin{figure}[h]
    \centering
    \includegraphics[width=\textwidth]{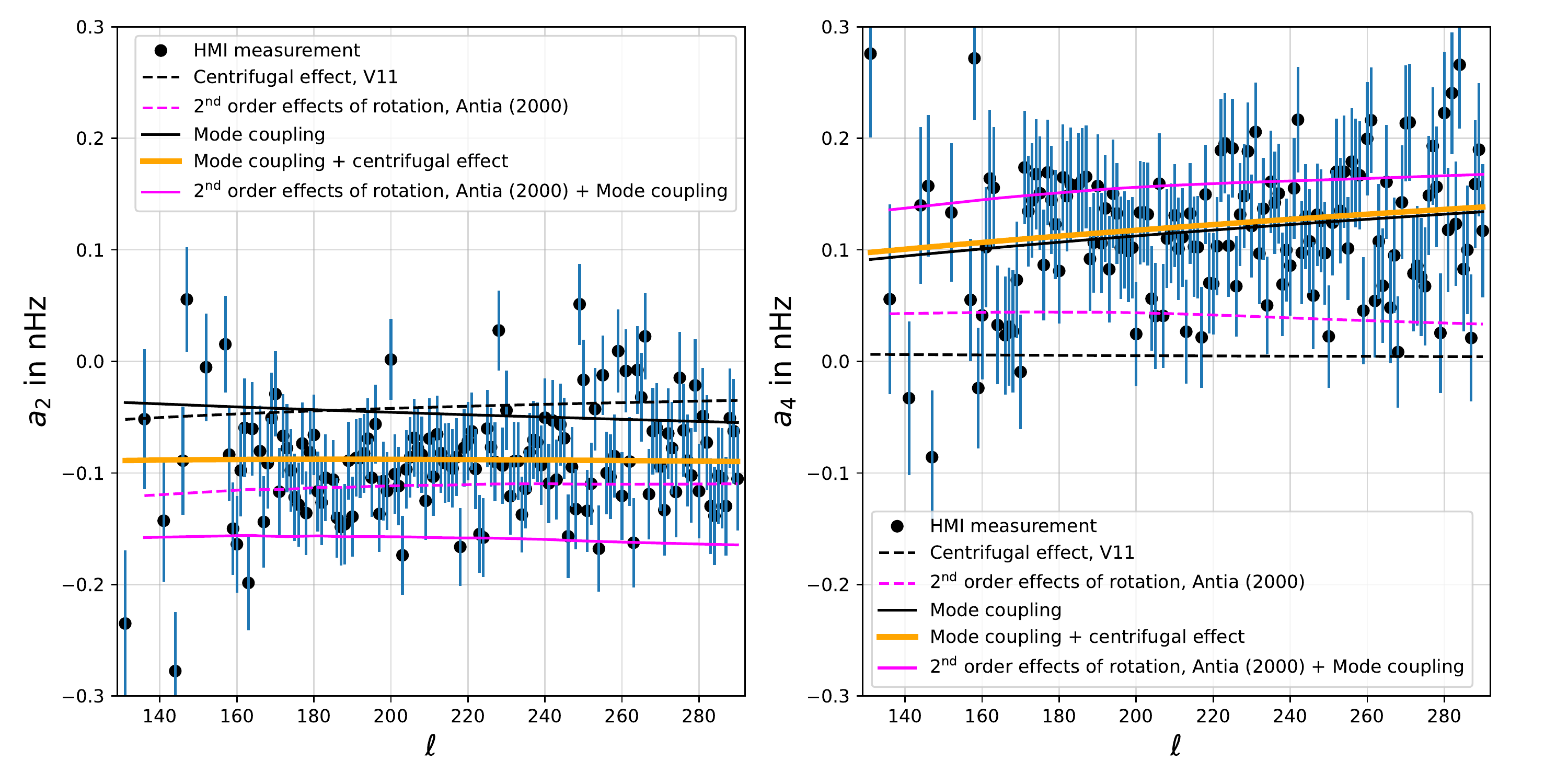}
    \caption{Same as Figures~1(a) and (b) in \cite{vorontsov11} but 
    using a numerical eigenvalue solver to estimate mode coupling. 
    For each subplot, the \textit{black-solid} line shows the 
    contribution from mode coupling (QDPT) and the \textit{black-dashed} 
    line marks the contribution from centrifugal effects. 
    The \textit{red} line denotes the combined contribution from 
    mode coupling and centrifugal effects. $a_2$ and $a_4$ HMI 
    measurements (from the 360-day period between 2010-04-30 and 
    2011-04-25, which coincides with the solar minimum) are marked as    \textit{black} dots. The \textit{pink} lines denote second order effects as presented in \cite{Antia2000}. For the \textit{pink} lines, we have used the same limits of angular degree $\ell$ in our plots as used by \cite{Antia2000}.}\label{fig:compare_evenacoeff_V11_minima}
    \centering
    \includegraphics[width=\textwidth]{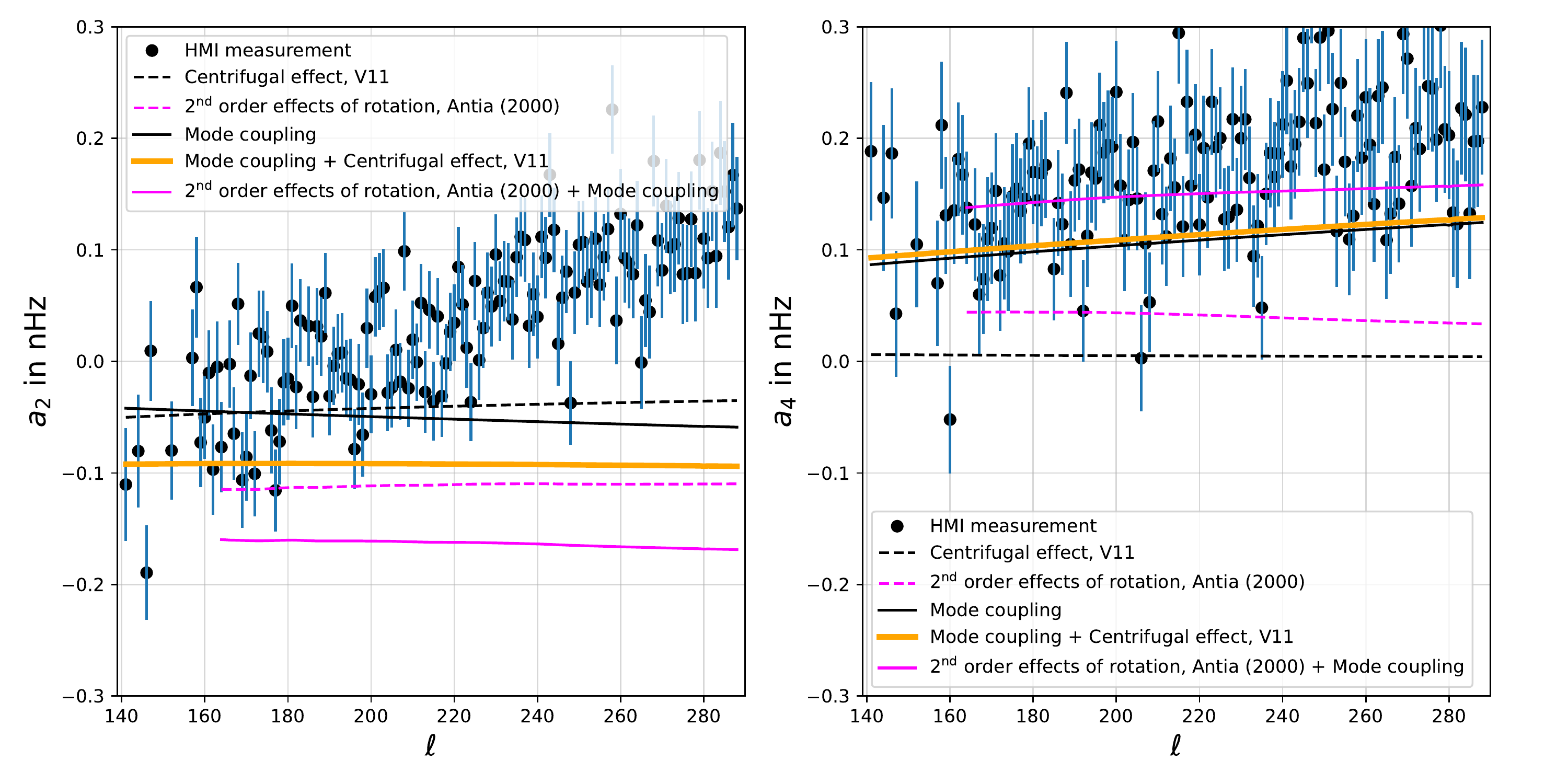}
    \caption{Same as Fig.~\ref{fig:compare_evenacoeff_V11_minima} but for the 360-day period of solar maxima between 2014-04-09 and 2015-04-04.}\label{fig:compare_evenacoeff_V11_maxima}
\end{figure}
V11 developed an asymptotic description of mode coupling for high
angular degrees and presented its effects on even $a$-coefficients.
In our study, solving the eigenvalue problem not only gives us the
odd $a$-coefficients, but also numerical estimates of even
$a$-coefficients. Therefore, for the sake of completeness, we also
tally our results, which were computed using a numerical eigenvalue solver, with those of V11. To elaborate further on how we obtained our estimates of $a_2^{n\ell}$ and $a_4^{n\ell}$, the following steps may be considered: (A) using Eqn.~(\ref{eqn:supmat_in_c}), we construct the supermatrix $\sfZ_{k'k}$ using the differential rotation profile for the chosen period (solar minima or maxima as discussed below), (B) since we are interested in considering mode-coupling, we carry out an eigenvalue problem to estimate the frequency splittings $\delta {}_{n} \omega_{\ell m}$, and (C) using Eqn.~(\ref{eqn:a_def}), we estimate the even $a$-coefficients for $j=2,4$.

Figs.~1(a) \& (b) in V11 compare
contributions of mode coupling (using a semi-analytic treatment) and
centrifugal effects on even coefficients $a_2$ and $a_4$ for the
$f$-mode. Reproducing the calculations of V11, the distortion
of solar surface due to centrifugal effects and the resulting
change in gravitational potential are written as 
\begin{equation}
    R(\theta) = R_\odot \left[ 1 + \epsilon_2 P_2(\cos\theta)
                            + \epsilon_4 P_4(\cos\theta)\right]
    \label{eqn:radial-distortion}
\end{equation}
\begin{equation}
    \Psi = -\frac{GM_\odot}{r} \left[ 
        1 - \left(\frac{R_\odot}{r} \right)^3 J_2 P_2(\cos\theta)
          - \left(\frac{R_\odot}{r} \right)^5 J_4 P_4(\cos\theta)
        \right],
\end{equation}
where $\epsilon_2,\, \epsilon_4$ are the oblateness coefficients,
$J_2,\, J_4$ are gravitational moments and $P_2,\, P_4$ are
Legendre polynomials. V11 showed that, the even-ordered 
$a-$coefficients can be written as 
\begin{equation}
    \delta a_2 = \left[ \frac{5}{2}\epsilon_2 + 3 J_2 - 
                        \frac{10}{3}(\epsilon_4 + J_4) \right]
                \left( \frac{\partial\omega}
                            {\partial\ell} \right)_n
    \label{eqn:centrifugal-a2}
\end{equation}
\begin{equation}
    \delta a_4 = \frac{3}{8}
        \left( 7\epsilon_4 + 10 J_4 \right)
                \left( \frac{\partial\omega}
                            {\partial\ell} \right)_n
    \label{eqn:centrifugal-a4}
\end{equation}
Using the measurements of $J_2,\, J_4$ from 
\cite{Roxburgh-2001-AnA}, V11 showed that 
the corrections to the even-ordered $a-$coefficients
can be written as 
\begin{equation}
    \delta a_2 = -1.20\times 10^{-5}  
    \left( \frac{\partial\omega}
                {\partial\ell} \right)_n,
    \qquad 
    \delta a_4 = -1.47\times 10^{-6}  
    \left( \frac{\partial\omega}
                {\partial\ell} \right)_n.
\end{equation}
From observed data, $\partial\omega/\partial\ell$
can be computed to be ${}_{n}\omega_{\ell+1} - 
{}_{n}\omega_{\ell}$. These calculations are shown in 
Fig.~\ref{fig:compare_evenacoeff_V11_minima}. V11 used one year of SOHO/MDI measurements over the solar
minimum to diminish contributions from magnetism. We carry out the
same analysis using results from an eigenvalue solver, but using
measurements from SDO/HMI during a 360-day period over a solar
minimum (2010-04-30 to 2011-04-25). From
Fig.~\ref{fig:compare_evenacoeff_V11_minima} we conclude that, in agreement
with V11, mode coupling and centrifugal effects together provide an
adequate fit to the observed $a_2$ coefficients. Mode coupling also
overwhelmingly dominates $a_4$ measurements, providing a good fit to
the data. However, for $a_2$, V11 observed mode coupling to dominate
over centrifugal effects beyond $\ell = 200$, whereas we find this to
happen beyond $\ell \sim 170$. We also plot the second-order effect of rotation as in \cite{Antia2000} in the \textit{dashed magenta} line and the combined contribution from mode-coupling and these second order effects in the \textit{solid magneta} line. The second-order correction in \cite{Antia2000} is a more complete treatment that the asymptotic calculation of centrifugal effects in V11. For both $a_2^{n\ell}$ and $a_4^{n\ell}$, just mode-coupling added with V11's asymptotic centrifugal effects seems to explain the HMI data at solar minima, accounting for a more rigorous second-order correction due to rotation clearly grazes the outer envelop of the \textit{black dots}. This suggests that there are other structure perturbations that need to be accounted for such as sound-speed anomaly or a weak background solar minima magnetic field.

Fig.~\ref{fig:compare_evenacoeff_V11_maxima} is plotted in the same spirit as Fig.~\ref{fig:compare_evenacoeff_V11_minima} but for the 360-day solar minima period between 2014-04-09 and 2015-04-04. Neither $a_2^{n\ell}$ nor $a_4^{n\ell}$ are explain by the effects from rotation alone. Since solar maxima has significantly more magnetic activity, this departure in the plotted even $a$-coefficients could be attributed to strengthened magnetic fields. According to selection rules of $a$-coefficient kernels for Lorentz-stresses \citep[see][]{sbdas20}, these departures in even $a$-coefficients may be using to infer the following magnetic quantities: $B_r^2, (B_{\theta}^2 + B_{\phi}^2), (B_{\theta}^2 - B_{\phi}^2)$ and $B_r \, B_{\theta}$.

\begin{figure}
    \centering
    \includegraphics[width=0.5\textwidth]{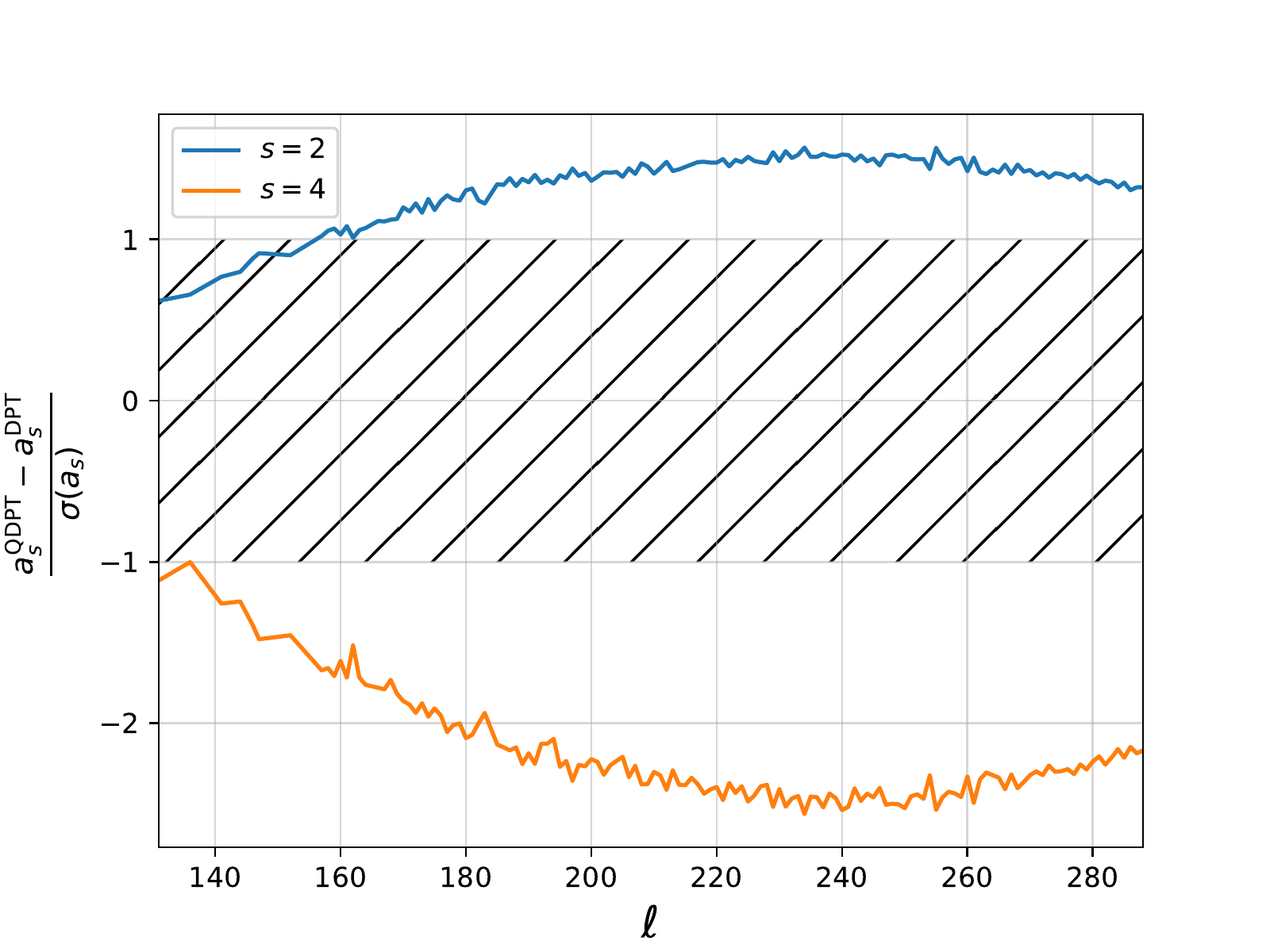}
    \caption{Departure of QDPT-estimated $a_2^{0\ell}$ and 
    $a_4^{0\ell}$ coefficients from the DPT estimates. Differences
    are scaled with respect to the standard deviations of the
    respective measurements. Statistically significant differences of
    magnitudes $> 1\sigma$ lie outside the \textit{black} hatched
    area. Forward calculations were performed using inverted rotation
    profiles available on JSOC, which are associated with HMI data
    measured during the 72-day period between 2014-04-09 and
    2014-06-20 which corresponds to a solar maxima.}
    \label{fig:a2_a4_QDPT_DPT}
\end{figure}

Measurements during solar maxima are expected to possess stronger signatures 
of magnetic fields as compared to the solar minima. Therefore, we also 
provide an estimate of the even $a$-coefficients (scaled by the measured
360-day uncertainties) using JSOC rotation profiles from the solar 
maximum between 2014-04-09 and 2015-04-09. We carried out full eigenvalue
solutions using $s_{\mathrm{max}}=19$ for all the modes observed by HMI. 
Only the $f$-modes were found to have predicted $a_2$ and $a_4$ which were 
larger than the uncertainties from a one-year period. The result for the modes 
resolved by HMI for $n=0$ is shown in Fig.~\ref{fig:a2_a4_QDPT_DPT}. While all 
measured modes have $a_4/\sigma(a_4) > 1$, only $\ell > 160$ modes have 
$a_2/\sigma(a_2) > 1$. The SNR for $a_4$ is about twice as large as $a_2$. If the same calculation is repeated for 360-day measurements instead of the currently used 72-day measurements, the SNR would be scaled up by $\sqrt{5}$.

\section{Discussion} \label{sec:discussion}
Measurement of differential rotation is one of the triumphs of helioseismology.
Traditional methods of inference have used frequency splittings $\delta
\omega_{n \ell}^m$ for global modes of oscillations under the assumption that
modes are self-coupled. While this is acceptable for most modes in observed
power spectra, this approximation worsens for high angular degrees in the $f$
and $p_1$ branches. \cite{Kashyap21} investigated the applicability of the
isolated-multiplet approximation across all the modes resolved by HMI up to
$\ell = 300$. They found deviations between $\delta \omega^{m,
\mathrm{QDPT}}_{n \ell}$ and $\delta \omega^{m, \mathrm{DPT}}_{n \ell}$ beyond
$\ell = 160$ of the order of $\approx \pm 2 \, \sigma(\delta \omega_{n \ell})$,
where $\sigma(\delta \omega_{n \ell})$ are the uncertainties in the measured
frequency splittings. The frequency splittings may be expressed in terms of odd
and even $a$-coefficients, which constrain the equatorially symmetric part of
differential rotation and axisymmetric structure perturbations, respectively.
This motivates the current study which is aimed at: (i) investigating the
changes in predicted odd and even $a$-coefficients via an eigenvalue treatment
of the operator $\sfZ$, (ii) finding corrections to differential rotation when 
considering mode coupling as compared to isolated multiplets, and (iii)
comparing the expected second-order changes in the even $a$-coefficients
computed from the exact eigenvalue solver as compared to asymptotic analysis
from previous studies, such as V11.

In this study, we adopt the QDPT approach to modeling the frequency splittings due to differential rotation. To the best of our knowledge, all studies that use $a$-coefficients to infer differential rotation assume that multiplets are isolated. We carry out forward modeling as well as an inverse problem by constructing the full supermatrix $\sfZ$ that accounts for coupling, with all neighbouring multiplets adhering to selections rules imposed by Wigner $3-j$~symbols. Section~8(c) in LR92 discusses the vanishingly small effect of QDPT for low angular-degree modes, quantified by the coupling-strength coefficient. Using an asymptotic analytical treatment, V11 proposed the maximum correction in frequency splittings as a function of $\ell$ and mentioned that the effect of mode coupling on oscillation frequencies might be observable towards high $\ell$. Our study demonstrates that the mode-coupling-induced corrections in odd $a$-coefficients introduce a relatively tiny change in differential rotation or torsional oscillations. We present this in Section~\ref{sec: DPT_vs_hyb} by performing 1.5D inversions for 22 years of MDI and HMI data. The corrections show a systematic variation over solar cycles, although the correction amplitudes are minuscule. This implies that accounting for mode coupling when inferring differential rotation solely from odd $a$-coefficients is insignificant. In Section~\ref{sec: even_acoeffs}, we also compare the results of our predicted even $a$-coefficients arising from mode coupling with those presented in the semi-analytic treatment of V11 and find them to be consistent with only very minor differences. Therefore, we believe that future studies accounting for even $a$-coefficients due to mode coupling can reliably use the corrections suggested by V11 instead of the more computationally expensive method performed here.

Apart from the correction in torsional oscillation (which is weak but has a systematic variation over solar cycles), we confirm that considering the contribution of differential rotation to even splitting coefficients due to mode coupling is statistically significant. Fig.~(\ref{fig:a2_a4_QDPT_DPT}) shows that this effect in $a_2^{n \ell}$ and $a_4^{n \ell}$ is as large as $1.5 \, \sigma$ and $2.5 \, \sigma$, respectively. The effect of mode coupling on $a_2^{n \ell}$ coefficients was accounted for in a comprehensive study by \cite{chatterjee-antia-2009} to put limits on flow velocities in giant cells. However, most studies that use even $a$-coefficients to infer global magnetism have not accounted for this correction. For instance, \cite{Antia2000} calculated estimates of magnetic field using forward modelling, after considering distortion due to second-order corrections of rotation on even $a$-coefficients. They predicted a magnetic field of strength 20~kG (or an equivalent acoustic perturbation) located 30~Mm below the solar surface using even coefficients $a_2^{n \ell}$ and $a_4^{n \ell}$. \cite{Baldner2009, baldner10} too carried out inversions for magnetic field strengths without accounting for mode coupling. Therefore, improving such prior estimates when the effect of QDPT is considered remains an important area in which to make progress. 

\acknowledgments
S.B.D and S.G.K thank Jesper Schou (Max Planck Institute for Solar System Research) for numerous insights and Pritwiraj Moulik (Princeton University) for kindly providing the B-splines package as well as multiple discussions on inversion methodology.

\bibliography{references}{}
\bibliographystyle{aasjournal}

\begin{appendix}
\section{Noisy $s=7$ HMI measurements}
\begin{figure}[h]
\centering
\includegraphics[width=0.9\textwidth]{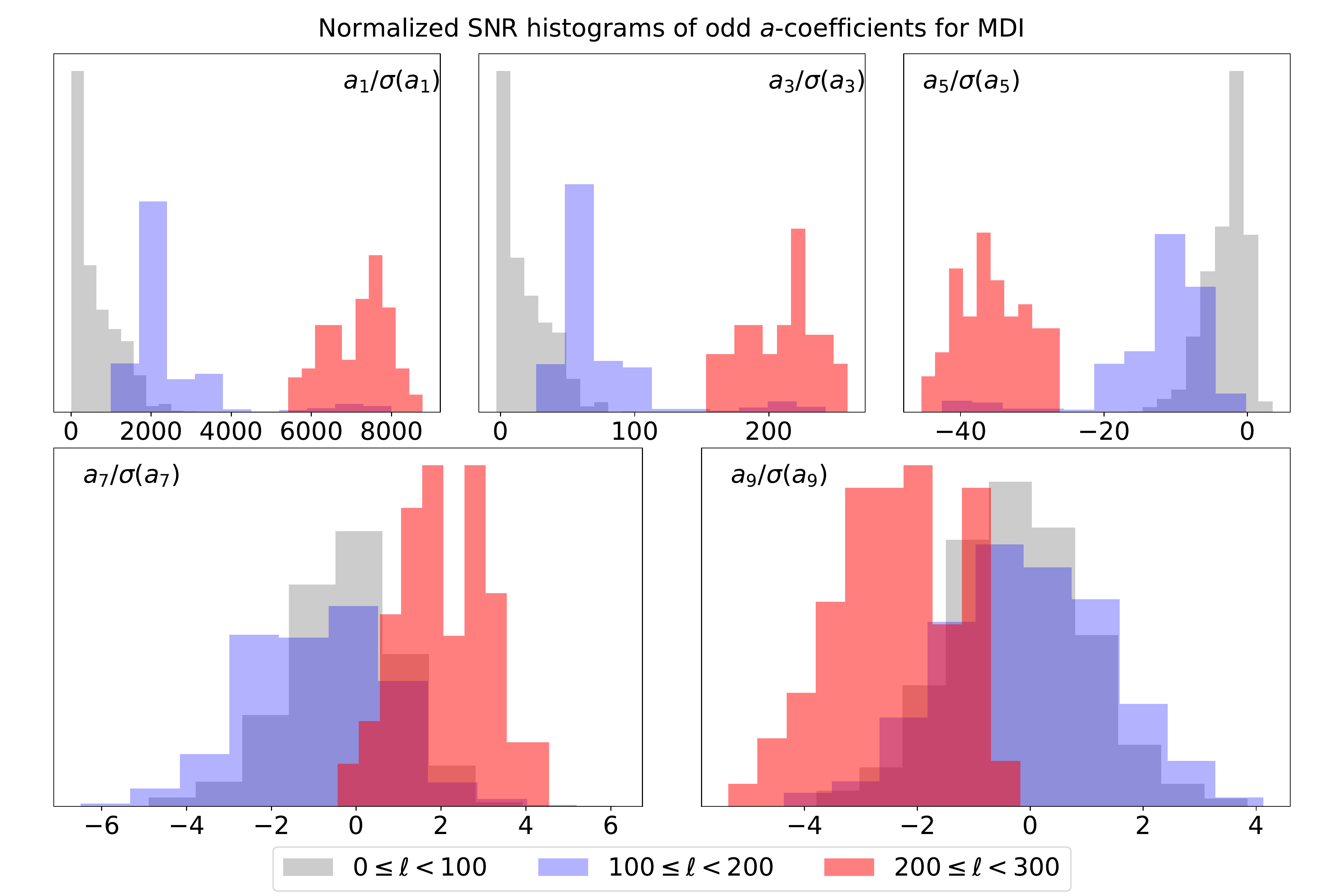}
\bigbreak
\includegraphics[width=0.9\textwidth]{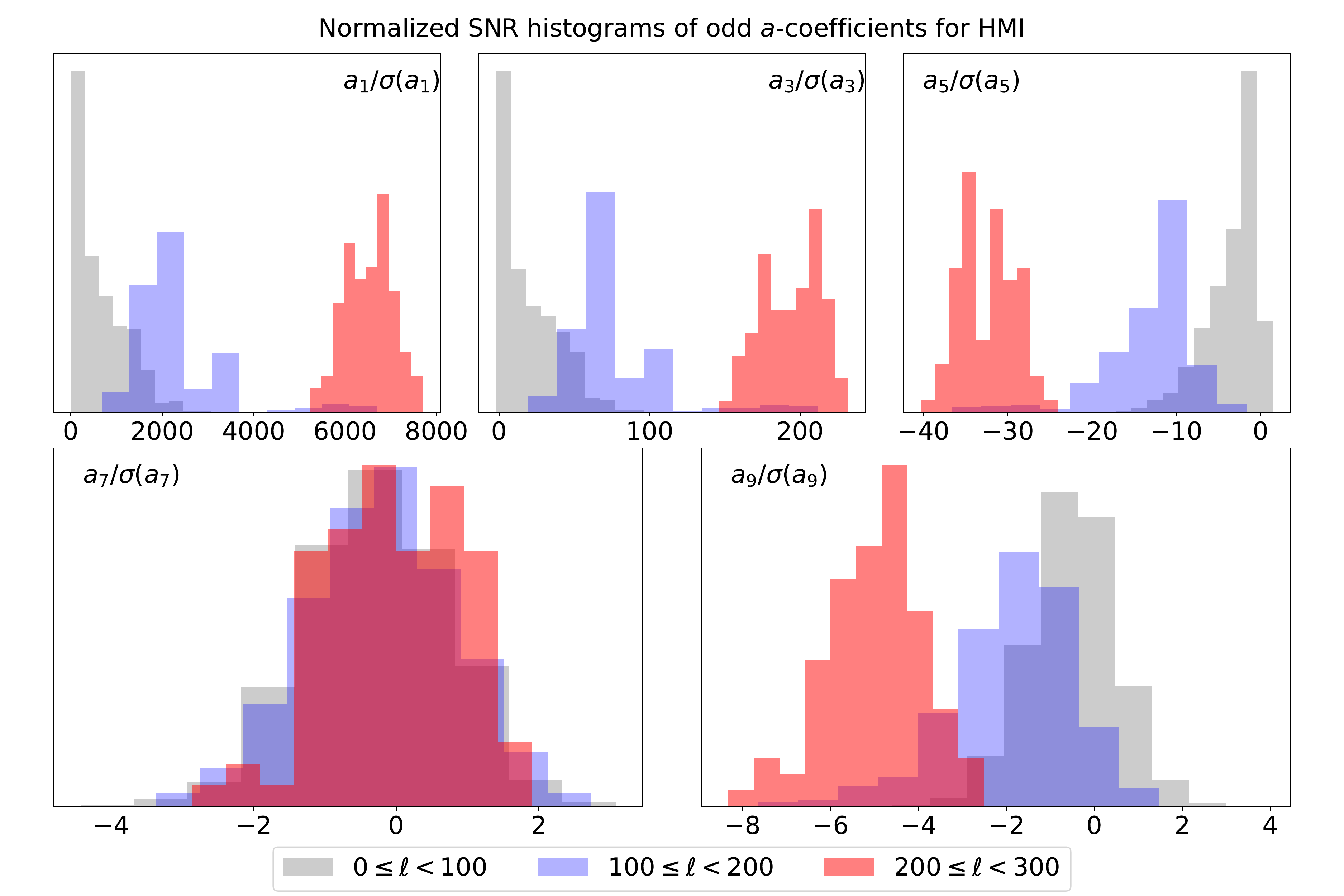}
\caption{Normalized histograms for signal-to-noise of odd $a$-coefficients $a_1 - a_9$. The \textit{gray}, \textit{blue} and \textit{red} colors indicate modes with low, intermediate and high angular degree $\ell$. Histograms in the \textit{top} panel were constructed from MDI data measured during the 72-day period between 1996-07-12 and 1996-09-22. Histograms in the \textit{bottom} panel were constructed from HMI data measured during the 72-day period between 2010-04-30 and 2010-07-11. Both of these periods coincides with solar minima.}
\label{fig:SNR_MDI_HMI}
\end{figure}

\begin{figure}[h]
\centering
\includegraphics[width=\textwidth]{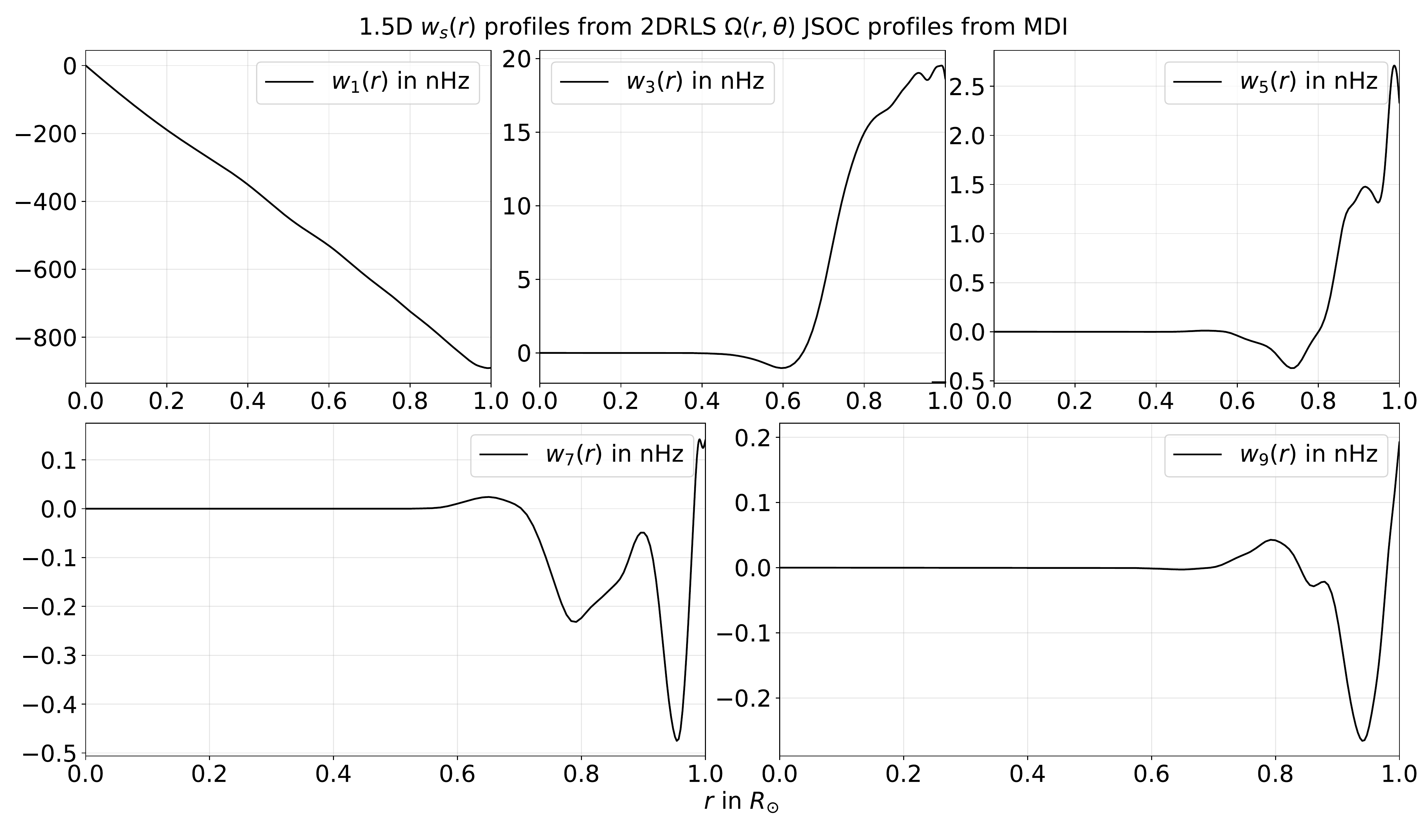}
\bigbreak
\includegraphics[width=\textwidth]{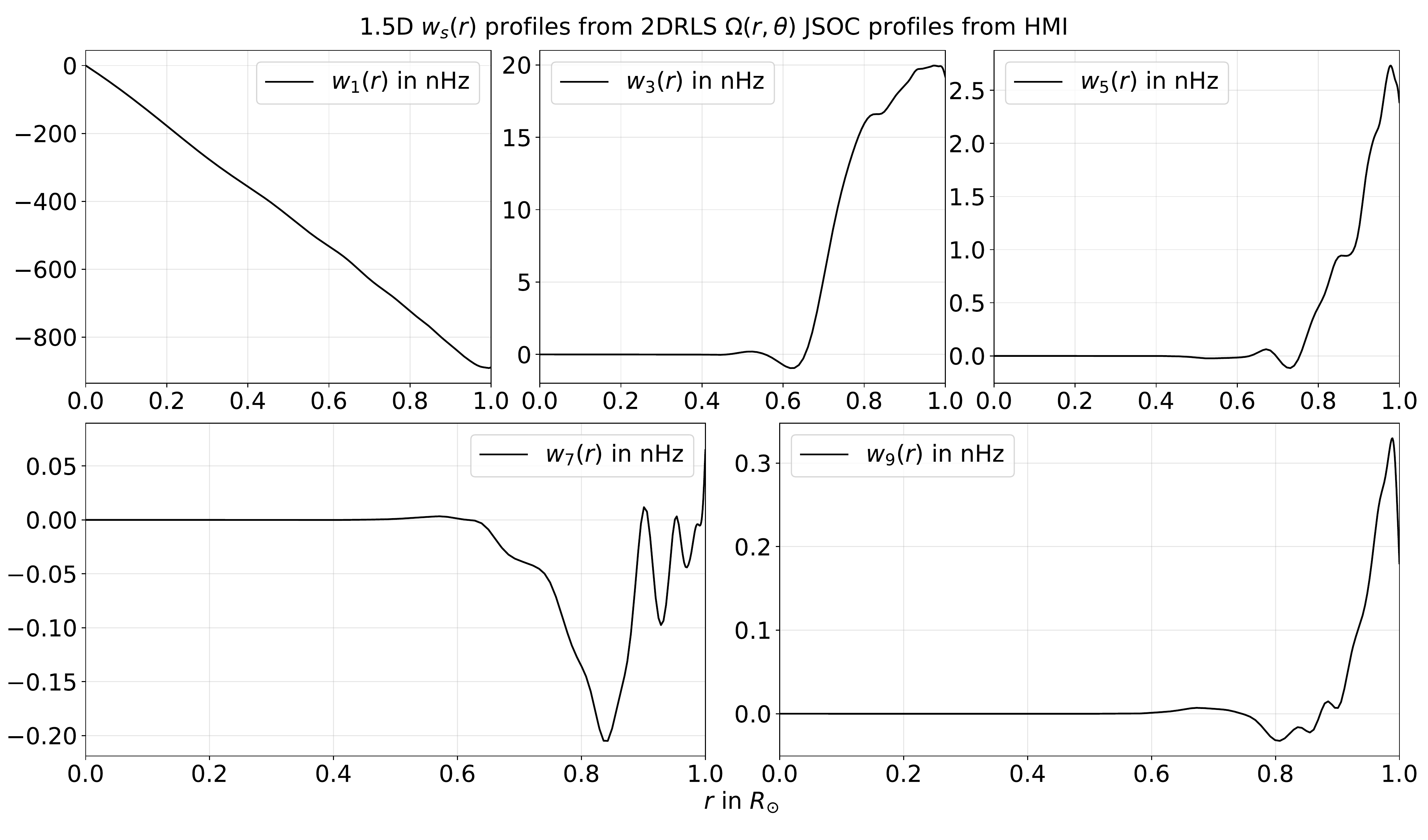}
\caption{1.5D counterparts of the 2D-RLS rotation profiles obtain from JSOC during two different solar minima phases. The profiles in the \textit{top} panel correspond to MDI measurements between 1996-07-12 and 1996-09-22 and those in the \textit{bottom} panel correspond to HMI measurements between 2010-04-30 and 2010-07-11.}
\label{fig:wsr_MDI_HMI}
\end{figure}

Inference of rotation as a function of depth depends critically on the signal-to-noise (SNR) of the measured $a$-coefficients of modes sensitive to perturbations at that depth. High (low) angular-degree modes have shallow (deep) lower turning points and are sensitive to shallow (deep) perturbations. Therefore, it is necessary to measure the SNR of the coupled modes. In doing so, we found that some of the 72-day HMI measurements have poor SNR for $a_7^{n \ell}$. This is reflected in the considerably wiggly $w_7(r)$ found from decomposing the 2D-RLS results on JSOC into 1.5D rotation profiles. Fig.~\ref{fig:SNR_MDI_HMI} shows normalized histograms for $a$-coefficients measured during phases of solar minima by MDI and HMI in three ranges of angular degrees: (a) low $\ell$ modes in \textit{gray} where $0 \leq \ell < 100$, (b) intermediate $\ell$ modes in \textit{blue} where $100 \leq \ell < 200$, and, (c) high $\ell$ modes in \textit{red} where $200 \leq \ell < 300$. For $a_1^{n \ell}, a_3^{n \ell}, a_5^{n \ell}$ and $a_9^{n \ell}$, the SNR for the shallow-sensitive modes is significantly larger than unity for both MDI and HMI. The SNR for only the HMI-measured $a_7^{n \ell}$, however, is largely contained within $\pm 1$. Consequently, as seen in Fig.~\ref{fig:wsr_MDI_HMI}, the inferred $w_s(r)$ profiles from MDI have the desired smoothness imposed by the regularization term. This is also true for $w_1(r), w_3(r), w_5(r)$ and $w_9(r)$ profiles from HMI. However, the $w_7(r)$ profile from HMI is unusually wiggly.

\section{Converting 2D rotation profiles to 1.5D} \label{sec: JSOC_to_1pt5D}
In the 1.5D inversions presented in Section~(\ref{sec: DPT_vs_hyb}), we fix $w_s(r)$ below $r_{\mathrm{th}}=0.9 R_{\odot}$ to the corresponding JSOC profiles (which were obtained via 2D RLS). This requires converting $\Omega(r, \theta)$ to its 1.5D equivalent $w_s(r)$. We follow the prescription of \cite{ritzwoller} which is outlined here for completeness and ease of reference of the reader. The rotation profile $\Omega(r, \theta)$ may be written in terms of 
Legendre polynomials $P_k$ as 
\begin{equation}
 \Omega(r, \theta) = \sum_{k=0, 2, 4...}\Omega_k(r) \, P_k(\cos\theta)
 \label{eqn:rot-legpoly}
\end{equation} 
 Further, we know that rotational velocity $\bfv_{\mathrm{rot}}(r, \theta) = \hat{\mathbf{r}} \times   \hat{\mathbf{z}} \, \Omega(r, \theta)$.
Using this and Eqn.~(\ref{eq: vrot_VSH}), we have
\begin{equation}
\mathrm{v}_{\mathrm{rot}} = 
r \sin\theta \sum_{k=0, 2, 4...} \Omega_k(r) \, P_k(\cos\theta)
= -\sum_{s=1, 3, 5, ...} \, w_s(r) \, \partial_\theta Y_{s0}
\end{equation}
Since our equations are written in terms of $w_s(r)$, we need to convert
$\Omega_k(r) \to w_s(r)$. To do this, we may project $\mathrm{v}_{\mathrm{rot}}$ onto the basis of $P_k$, i.e.,
\begin{equation}
r\,\Omega_k(r) = - \frac{2}{2k+1}\int
\frac{1}{\sin\theta}\left( \sum_{s=1, 3, 5, ...} w_s(r)\, \partial_\theta Y_{s0}(\theta)
\right) P_k(\cos\theta) \sin\theta \, \rmd\theta  \, ,
\end{equation}
which gives us the following matrix equation
\begin{equation}
r \, \Omega_k(r) = \sum_s \alpha_{ks} \, w_s(r)
\end{equation}
Finally, $w_s(r)$ can be obtained by $w_s(r) = \boldsymbol{\alpha}^{-1} \, \Omega_k(r)$. To obtain upto $w_9(r)$, we use the following matrix inverse
\begin{equation}
\boldsymbol{\alpha}^{-1} = 2\sqrt{\pi}\left(
\begin{array}{cccccc}
 1/\sqrt{3} & -1/5 \sqrt{3} & 0 & 0 & 0 & 0 \vspace{0.2cm}\\
 0 & 1/5 \sqrt{7} & -1/9 \sqrt{7} & 0 & 0 & 0 \vspace{0.2cm}\\
 0 & 0 & 1/9 \sqrt{11} & -1/13 \sqrt{11} & 0 & 0 \vspace{0.2cm}\\
 0 & 0 & 0 & 1/13 \sqrt{15} & -1/17 \sqrt{15} & 0 \vspace{0.2cm}\\
 0 & 0 & 0 & 0 & 1/17 \sqrt{19} & -1/21 \sqrt{19} \vspace{0.2cm}\\
 0 & 0 & 0 & 0 & 0 & 1/21 \sqrt{23} \\
\end{array}
\right)
\end{equation}

\section{Accuracy and robustness of the numerical eigenvalue solver}\label{sec: eig_accuracy}
\begin{figure}
    \centering
    \includegraphics[width=\textwidth]{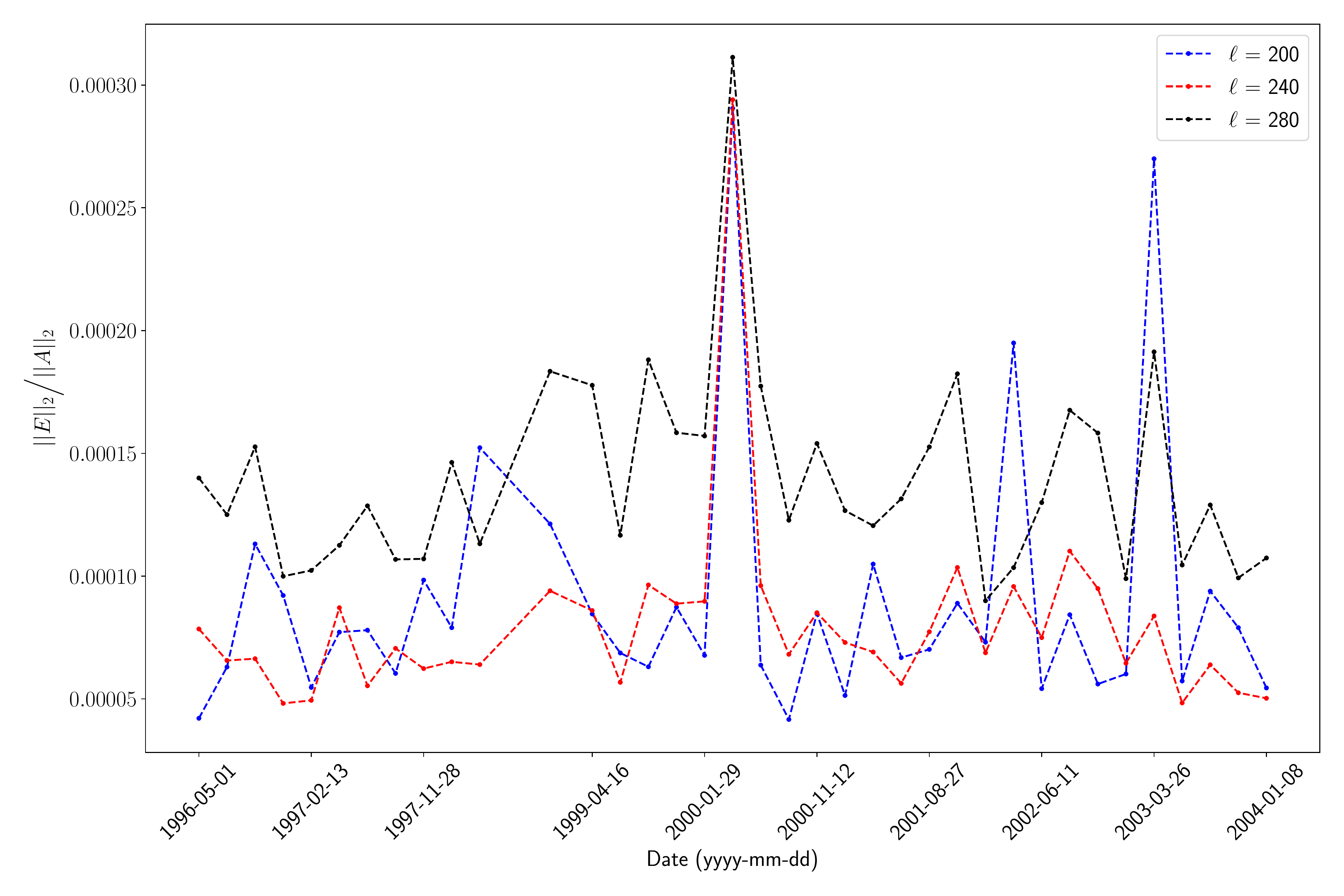}
    \caption{Ratio of the L2-norm of $E$, the difference between the hybrid and DPT supermatrices, and the L2-norm of the DPT supermatrix $A$. The \textit{blue, red} and \textit{black} dashed lines correspond to angular degrees $\ell = 200, 240, 280$, respectively.}
    \label{fig:E-by-A}
\end{figure}

We use the \texttt{linalg.eigh} module in Python's \texttt{numpy} package for computing eigenvalues. The \texttt{lialg.eigh} module is a wrapper for its LAPACK implementation of evaluating eigenvalues of real symmetric or complex Hermitian matrices. As mentioned in the documentation webpage of \href{https://numpy.org/doc/stable/reference/generated/numpy.linalg.eigh.html}{\texttt{numpy.linalg.eigh}}, for a symmetric and real matrix (as is the case for our supermatrix), the \href{https://www.intel.com/content/www/us/en/develop/documentation/onemkl-developer-reference-fortran/top/lapack-routines/lapack-least-squares-and-eigenvalue-problem/lapack-least-squares-eigenvalue-problem-driver/symmetric-eigenvalue-problems-lapack-driver/syevd.html}{\texttt{\_syevd}} routine is used for solving a real symmetric matrix using divide and conquer algorithm. Under ``Application Notes'' in the \texttt{\_syevd} webpage, the developers mention that ``The computed eigenvalues and eigenvectors are exact for a matrix A+E such that $||E||_2 = \mathcal{O}(\epsilon) \, ||A||_2$, where $\epsilon$ is the machine precision''. In our case, $A$ may be regarded as the DPT supermatrix and $E$ would then be the difference between the hybrid and DPT supermatrices. Note that the hybrid supermatrices are constructed from the final converged $c_s^{p, \mathrm{fit}}$ arrays (see Eqn.~[\ref{eqn: build_full_supmat}]) obtained from our inversion algorithm --- the coefficients we use to construct the final plots Figs.~\ref{fig: timeseries_lat} \& \ref{fig: timeseries_depth}.

For our problem, we compute the DPT supermatrix (which ignores coupling across multiplets) and the hybrid supermatrix (which accounts for coupling across multiplets). Since the coupling across multiplets induced due to differential rotation is weak --- as evidenced by the smallness of our corrections, it is reasonable to ask if the matrices themselves are \textit{different enough} for LAPACK's \texttt{\_syevd} algorithm to yield distinctly different eigenvalues. We can frame this question, in light of the Application Note provided by the LAPACK developers, as: ``Is the ratio of L2-norm of our $E$ matrix (difference between hybrid and DPT supermatrix) and the L2-norm of our DPT matrix $A$ significantly larger than the machine precision?'' We carry out our calculations in 64-bits precision, meaning $\mathcal{O}(\epsilon) \approx 10^{-16}$. So, to investigate the above question, we have calculated $||E||_2 \big/ ||A||_2$ across multiple years for three large angular degrees $(\ell = 200, 240, 280)$ where hybrid fitting is applicable. The results are presented in Fig.~\ref{fig:E-by-A}. We see that the ratio is consistently larger than $\mathcal{O}(10^{-6})$ which is atleast $10^{10}$ times larger than the order of machine precision $\mathcal{O}(\epsilon)$. Therefore, our DPT supermatrix $A$ and the corresponding hybrid supermatrix $A+E$ are different enough to yield distinctly different eigenvalues. This shows that the eigenvalue routine used in Eqn.~(14) is accurate enough to yield eigenvalues which are not garbled by machine errors for the level of difference between DPT and hybrid supermatrices. 

Eqn.~(15) shows the cost function and the inversion involves fitting the model parameters $c_s^{p, \mathrm{fit}}$ while trying to minimize this cost function. Since our problem involves an eigenvalue operation, this is non-linear and so we adopt the standard Newton's method \citep{Tarantola-1987-iptm}. This involves starting from a guess solution (which is close enough to the \textit{true} solution than minimizes the cost function) and iteratively stepping towards the minima by computing a gradient vector and Hessian tensor at each updated model parameter. These are commonplace in machine learning community and Google's \href{https://jax.readthedocs.io/en/latest/}{\texttt{jax}} package has emerged as an efficient tool for the same \citep{jax2018github}. In our study, we have used the automatic differentiation routines \texttt{jax.grad} to compute gradients and the routines \texttt{jax.jacfwd} and \texttt{jax.jacrev} to compute the Hessian. Note that \texttt{jax} uses 32-bits precision by default but we have used the additional switch \texttt{config.update('jax\_enable\_x64', True)} to use 64-bit machine precision in our calculations.

\section{Validation of inversion}\label{sec: inversion_validation}

\begin{figure}
    \centering
     \includegraphics[width=\textwidth]{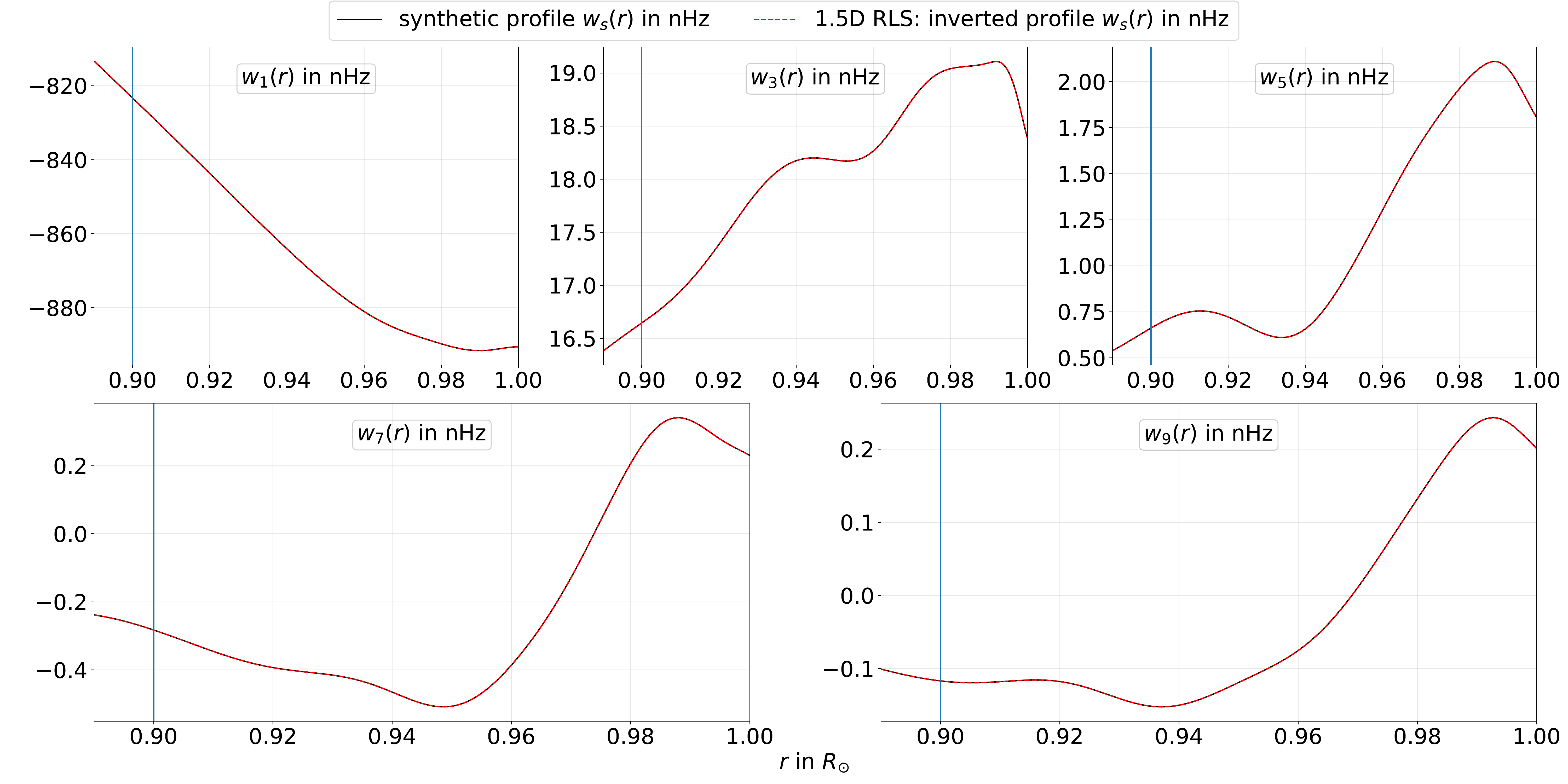}
     \caption{Validation of noise-free non-linear inversion. We construct synthetic dataset with $a-$coefficients
     generated using Eqn.~(\ref{eqn: model_DPT_hybrid}), shown by the \textit{solid black} curve. We carry out non-linear inversions for the $w_s(r)$ profiles using these 
     synthetically generated $a-$coefficients. The inverted profiles are shown by the \textit{dashed red} lines.}
    \label{fig:no-noise-synth-inversion}
\end{figure}

\begin{figure}
    \centering
    \includegraphics[width=\textwidth]{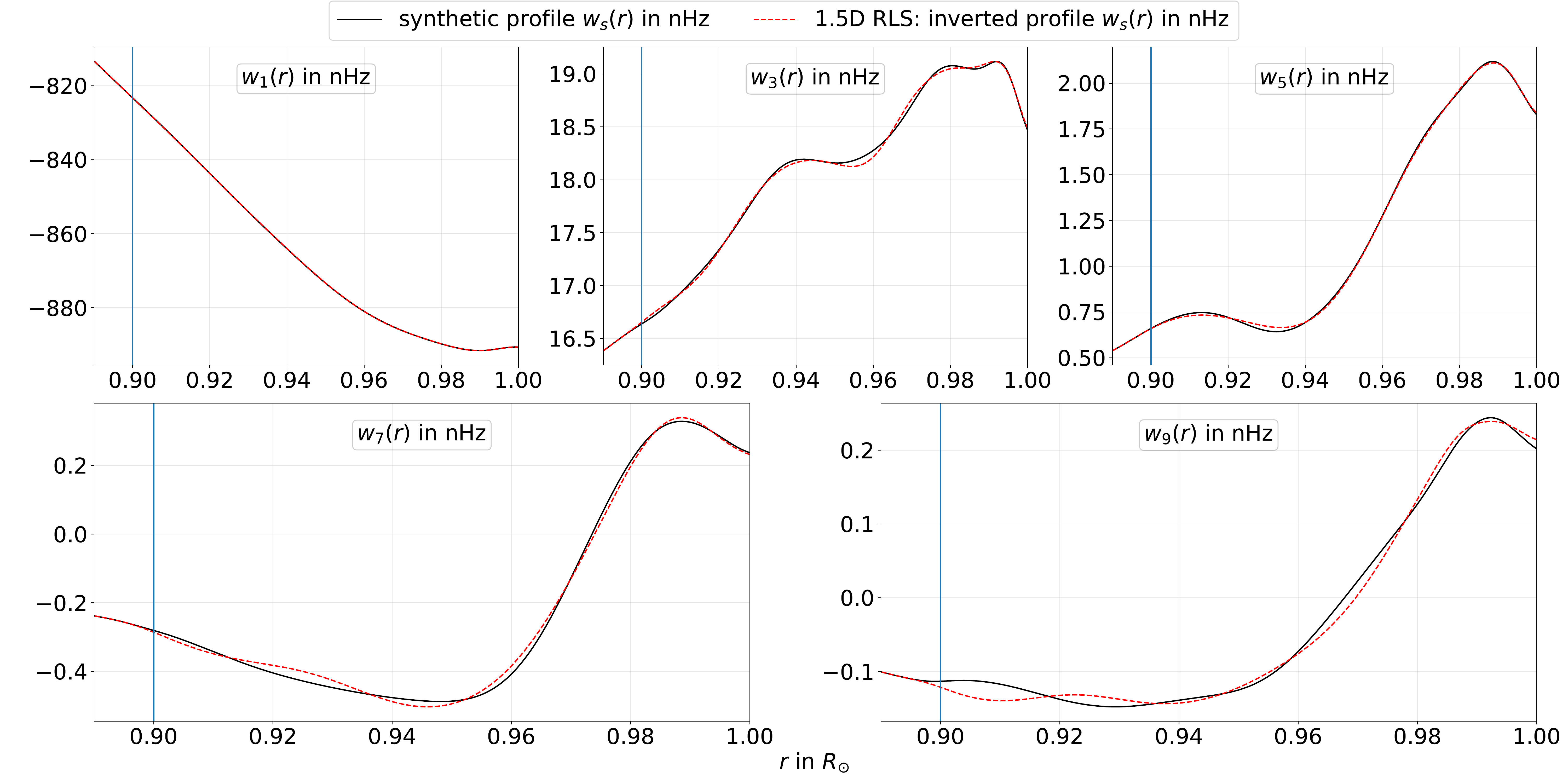}
    \caption{Validation of inversion with noise. As in Fig.~\ref{fig:no-noise-synth-inversion}, the \textit{solid black} line indicates the synthetic $w_s(r)$ profiles to which we add noise before carrying out a non-linear inversion to infer the \textit{dashed red} line.}
    \label{fig:noisy-synth-inversion}
    \bigbreak
    \includegraphics[width=\textwidth]{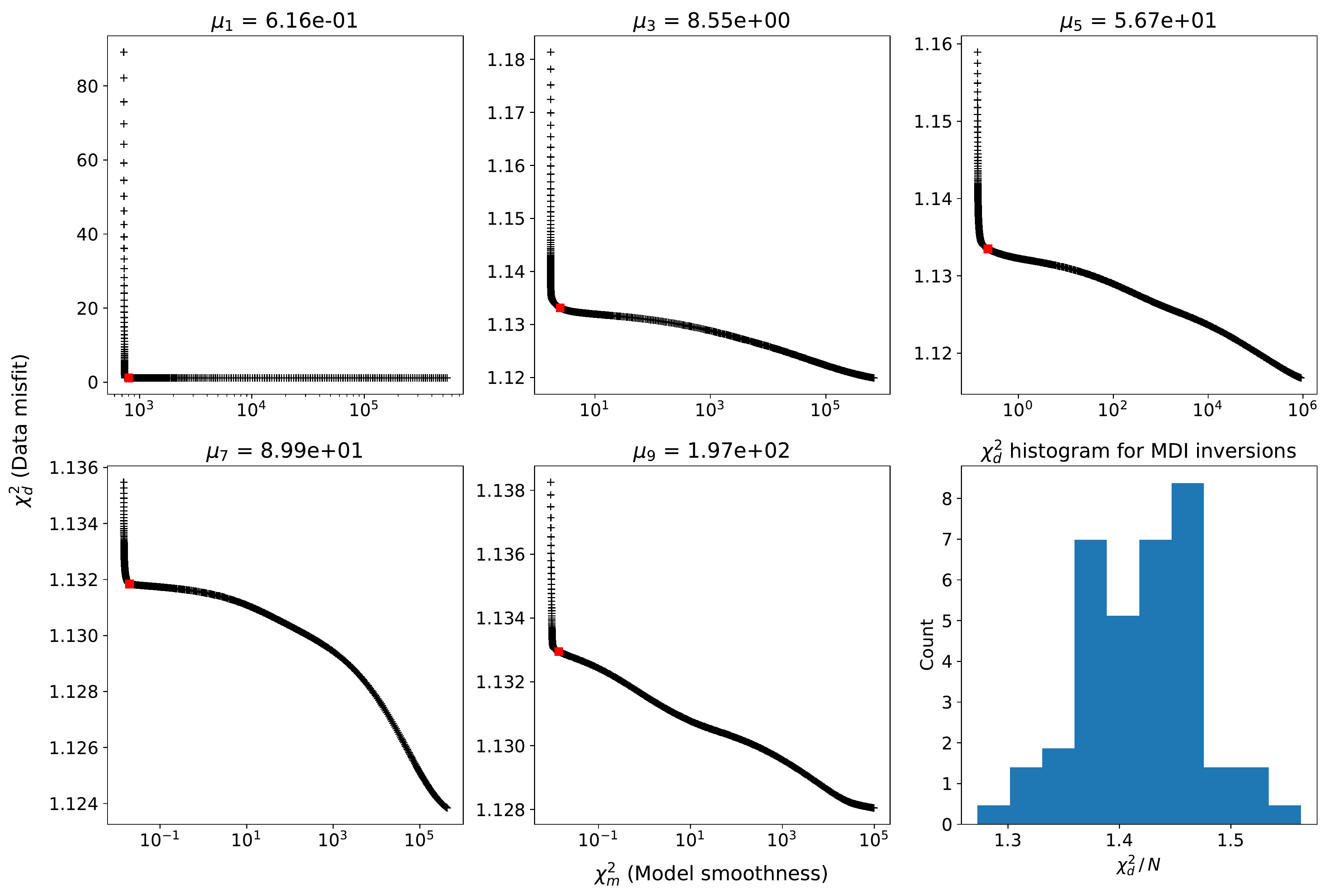}
    \caption{The first five panels show L-curves for the different angular degrees $s = \{1,3,5,7,9\}$. These plots are representative of L-curves corresponding to a typical 72-day MDI or HMI inversion. The red marker in each plot shows our chosen \textit{knee} of the curve and the corresponding value of $\mu_s$ is mentioned in the title of each subplot. The last panel in the lower-right, shows a histogram (normalized to yield unit area) of $\chi^2_d/N$ for all MDI inversions.}
    \label{fig:Lcurve}
\end{figure}

This section demonstrates the validity of the non-linear hybrid inversions using synthetic, yet realistic rotation profiles. To do so, we choose a reference profile $\mathbf{v}_{\mathrm{rot}}^{\mathrm{ref}}(r,\theta)$ available through the JSOC pipeline corresponding to the 72-day MDI $a$-coefficient measurements starting on $12^{\mathrm{th}}$ July, 1996. We then use Eqns.~(\ref{eqn:spline_decomposition}), (\ref{eqn: build_full_supmat}) \& (\ref{eqn: model_DPT_hybrid}) to generate $a$-coefficients using the hybrid method, i.e., isolated multiplet treatment for DPT modes and full QDPT treatment for coupled modes. We first carry out an inversion using these noise-free synthetically generated $a$-coefficients. Fig.~\ref{fig:no-noise-synth-inversion} compares the spherical harmonic components of the inverted profile $\mathbf{v}_{\mathrm{rot}}^{\mathrm{inv}}$ with those of the synthetic profile $\mathbf{v}_{\mathrm{rot}}^{\mathrm{ref}}$ (see Eqn.~[\ref{eq: vrot_VSH}] for the spherical harmonic decomposition). The two profiles are seen to be exactly the same indicating the validity of noise-free inversion methodology. Similarly, inversions using data corrupted with synthetic noise (at 0.1 $\sigma$, where $\sigma$ represents the level of noise from observed $a$-coefficients) is shown in Fig.~\ref{fig:noisy-synth-inversion}. In this case, the $\mathbf{v}_{\mathrm{rot}}^{\mathrm{inv}}$ and $\mathbf{v}_{\mathrm{rot}}^{\mathrm{ref}}$ are within the expected errors with neither large fluctuations nor overly smoothened profiles (both of which are usually seen for improperly regularized inversions in the presence of noisy data). Hence, we deem the inversions to be successfully benchmarked using realistic rotation profiles. We have performed this 2-pronged test (first clean and then noisy inversion) over a variety of different profiles to validate the robustness of our inversion methodology.

The first five panels in Fig.~\ref{fig:Lcurve} show representative L-curves for determining regularization parameters $\mu_s$ corresponding to the different angular degrees $s= \{1, 3, 5, 7, 9\}$. The \textit{red} marker indicates the chosen optimal value of $\mu_s$ located at the \textit{knee} of the curve. This choice of regularization parameter represents an optimal balance between the data misfit (the degree to which theoretical predictions from our inferred profile matches the observed data) and the model smoothness (the degree of smoothness we impose in order to generate physically meaningful solutions and avoid unrealistic oscillatory behaviour of inferred profiles). The lower-right panel in Fig.~\ref{fig:Lcurve} shows a distribution of the data misfit (scaled by the total number of data points $N$) where the histogram is normalized to yield unit area. The distribution peaks around a scaled chi-square value of 1.4 which validates that our inversion is not over-fitting while producing inferred profiles which generate predictions having high-fidelity to the observed $a$-coefficients. The histogram is created using chi-square values from all MDI inversions.

\begin{figure}
    \centering
    \includegraphics[width=0.7\textwidth]{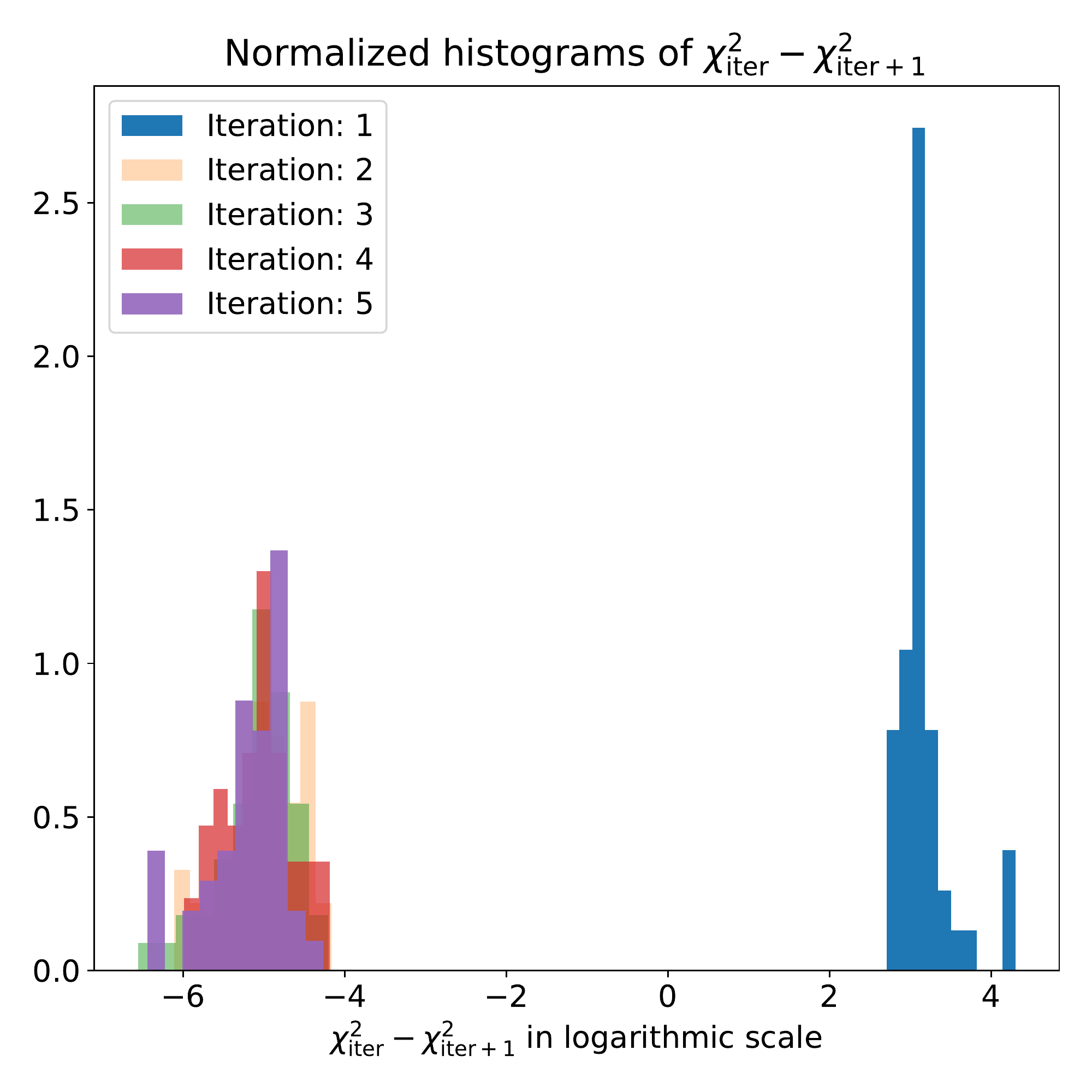}
    \caption{Histograms showing the change in $\chi^2$ value between successive iterations across all 72-day MDI inversions. Histograms are color-coded according to their iteration number, as shown in the figure legend. Each histogram is normalized to unit area.}
    \label{fig:chisq_convergence}
\end{figure}

Fig.~\ref{fig:chisq_convergence} is presented to demonstrate the convergence of our iterative non-linear Newton inversion. Each histogram shows the distribution of the change in total chi-square value between successive iterations. The histogram corresponding to the first iteration shows that, on an average, the total misfit drops by $\mathcal{O}(10^3)$ in the first iteration. Thereafter, the following iterations are seen to cluster around a much smaller change in misfit --- around 8 orders of magnitude smaller than the change in the first iteration. For the hybrid inversions, we start from the DPT profile which is obtained from carrying out a linear inversion under the isolated multiplet approximation. Therefore, Fig.~\ref{fig:chisq_convergence} shows that during the iterative inversions the rotation profiles undergo an initial non-trivial change from the DPT profiles, followed by negligible or insignificant changes in the following iterations. We run each of our inversions for five iterations to ensure the above convergence in misfit is achieved for all the cases.

\end{appendix}
\end{document}